\documentclass[aps,prl,reprint,groupedaddress,longbibliography]{revtex4-1}

\usepackage{graphicx}
\usepackage{color}
\pdfoutput=1

\begin{document}

\newcommand{\vet}[1]{\mathbf{#1}}
\newcommand{\vect}[1]{\mathbf{\textit{#1}}}
\newcommand{\an}[1]{{\color{blue} \bf{#1}}}
\newcommand{\lc}[1]{{\color{red} \bf{#1}}}



\title{Biaxial extensional viscous dissipation in sheets expansion formed by impact of drops of Newtonian and non-Newtonian fluids}


\email[]{christian.ligoure@umontpellier.fr, laurence.ramos@umontpellier.fr}

\author{Ameur Louhichi$^{1,2}$, Carole-Ann Charles$^{1}$, Ty Phou$^{1}$, Dimitris Vlassopoulos$^{2}$, Laurence Ramos$^{1}$ and Christian Ligoure$^{1}$}

\affiliation{
$^1$ Laboratoire Charles Coulomb (L2C), Universit\'e de Montpellier, CNRS, Montpellier, France\\
$^2$ Institute of Electronic Structure and Laser, FORTH, Heraklion 70013, Crete, Greece and Department of Materials Science and Technology, University of Crete, Heraklion 70013, Crete, Greece\\
}


\date{\today}

\begin{abstract}

We investigate freely expanding liquid sheets made of either simple Newtonian fluids or solutions of high molecular water-soluble polymer chains. A sheet is produced by the impact of a drop on a quartz plate covered with a thin layer of liquid nitrogen that suppresses shear viscous dissipation thanks to an inverse Leidenfrost effect. The sheet expands radially until reaching a maximum diameter and subsequently recedes. Experiments indicate the presence of two expansion regimes: the capillary regime, where the maximum expansion is controlled by surface tension forces and does not depend on the viscosity, and the viscous regime, where the expansion is reduced with increasing viscosity. In the viscous regime, the sheet expansion for polymeric samples is strongly enhanced as compared to that of Newtonian samples with comparable zero-shear viscosity. We show that data for Newtonian and non-Newtonian fluids collapse on a unique master curve where the maximum expansion factor is plotted against the relevant effective \textit{biaxial extensional} Ohnesorge number that depends on fluid density, surface tension and the biaxial extensional viscosity. For Newtonian fluids, this biaxial extensional viscosity is six times the shear viscosity.  By contrast, for the non-Newtonian fluids, a characteristic \textit{Weissenberg number}-dependent biaxial extensional viscosity is identified, which is in quantitative agreement with experimental and theoretical results reported in the literature for biaxial extensional flows of polymeric liquids.

\end{abstract}


\maketitle


\section {Introduction}

Over the last 15 years, the understanding of drop impact on solid targets has progressed considerably thanks to high-speed imaging methods~\cite{THORODDSEN:2008kl}, allowing one to observe in real time the fate of a drop upon impact under various experimental conditions and to probe a rich variety of phenomena, including dynamics of sheets in the expansion and receding regimes, spatio-temporal evolution of the thickness of the sheets~\cite{Vernay:2015eg, Wang:2017ju}, fingering instabilities~\cite{Yoon:2007ft, THORODDSEN:1998kk, Marmanis:1996ka}, fractures and production of satellite droplets~\cite{vanderMeer:2017bk, Josserand:2016jf, Bonn:2009ha}. Concerning the nature of the impacting drops, mainly Newtonian fluids of different viscosities have been investigated, although a few relevant studies with shear-thickening fluids~\cite{Boyer:2016gl}, shear-thinning fluids~\cite{Laan:2014hj, An:2012gc, CooperWhite:2002vt, Crooks:2000bt, Izbassarov:jq, Rozhkov:2006ic, Huh:2015jq}, yield stress fluids~\cite{Luu:2009jr}  or Maxwell fluids without shear-thinning~\cite{Arora:2016bu, Arora:2018ei}. Most studies have been performed on drops impacting a flat surface that can be smooth or rough \cite{Rioboo:2002gk}, horizontal or tilted  \cite{Moreira:2007gh}, hydrophobic or hydrophilic \cite{Lee:2010bt}. Often, surfaces have a very large size compared to that of the drop such that the entire spreading event occurs on the target~\cite{Josserand:2016jf, German:2009hp, Crooks:2000bt, Lee:2016ee, Izbassarov:jq, Ukiwe:2005jw, Roux:2004dx}, but targets of size comparable to that of the drops~\cite{Rozhkov:2004in, Rozhkov:2006ic, Villermaux:2011ff, Vernay:2015eg, Wang:2017ju} or drops impacting only partially a small target~\cite{Lejeune:2018da} have also been studied.

The complex interaction of a drop with a solid surface during drop collision may be removed or at least significantly reduced, by using  repellent surfaces, which avoid a direct contact between the liquid sheet and the solid target. Repellent surfaces include superhydrophobic surfaces~\cite{Richard2002}, hot plates above the Leidenfrost temperature~\cite{Wachters:1966vy, Biance2003} or sublimating surfaces~\cite{Antonini2013, Arora:2018ei}. Nevertheless, the fact that shear viscous dissipation can be neglected during the expansion of the sheet after impact does not mean that there is no viscous dissipation process. Indeed, \textit{biaxial extensional}~viscous dissipation is dominant in freely expanding sheets. Surprisingly this has never been documented to the best of our knowledge, except in a very recent paper~\cite{Pack:2019hn}, where the authors have attributed an inhibition of a drop-substrate contact during drop impact to a large increase of the extensional viscosity.

A possible reason for ignoring the biaxial extensional viscous dissipation is that for sheets expanding completely on a solid surface, viscous dissipation should be dominated by shear. For small targets however, both shear and biaxial extensional viscous dissipation processes may be relevant: this is the goal of a future publication. In the present paper, only biaxial extensional viscous dissipation is relevant as the sheet expands freely thanks to the inverse Leidenfrost effect discussed below. In this work, we investigate the expansion dynamics of free sheets of a viscoelastic thinning  fluid produced upon impacting a single drop on a repellent surface in inverse Leidenfrost conditions, and compare it to the respective response of Newtonian fluids. We demonstrate that accounting for the viscous dissipation due to biaxial extensional viscosity during the expansion of the sheet  is crucially important. We provide a simple approach to evaluate the biaxial extensional viscosity of thinning viscoelastic fluids, and finally propose a model to quantitatively account for the viscosity dependence of the maximum expansion of sheets. The paper is organized as follows. We first describe the materials and methods. We then show the shear rheological properties of the viscoelastic fluids of interest and their behavior upon impact on a repellent surface. Subsequently, we rationalize the results by accounting for the biaxial extensional viscosity and by means of a simple scaling model. The main conclusions are summarized in the last section.


\section {Materials and methods}
\subsection {Materials}

 We investigate solutions of polyethylene-oxide (PEO) of high molecular weight ($ M_\nu= 8000$ kDa) from  Sigma-Aldrich. Several samples with concentration $C$  between {$10^{-3} \rm{wt}\%$} and {$2 \rm{wt}\%$} are prepared by adding PEO powder, as received, to the appropriate volume of deionized pure water, or mixtures of water and glycerol ($20\rm{wt}\%$, $35\rm{wt}\%$ and $41.66\rm{wt}\%$ glycerol), and letting the solution under stirring at $T=25^{\circ}$C for 24 hours at least in the dark,   until complete dissolution. Note that, in order to enhance the visualization contrast, all PEO solutions are colored using a Nigrosin dye (From sigma-Aldrich) at concentration {$0.025 \rm{wt}\%$}. The surface tension of high molecular weight PEO solutions is independent of polymer concentration ($\gamma=62 \mathrm {mNm^{-1}}$)~\cite{Cao:1994kv}. Pendant drop experiments confirm that the addition of dye does not affect the surface tension of the final solutions. In addition, in order to compare the behavior of viscoelastic polymer solutions with that of simple fluids, we use two classes of Newtonian fluids, mixtures of water and glycerol and silicon oils. Silicon oils, with  shear viscosities from $5.2$ mPa.s to $1075$ mPa.s and an average surface tension of $20$ mN/m~\cite{ISI:A1987G694200038}, are purchased from Sigma Aldrich and used as received. Mixtures of water and glycerol with concentrations ranging from $22$ to $97.5$ $\rm{wt}\%$ glycerol are prepared, yielding  shear viscosities from $1.7$ mPa.s to $1910$ mPa.s ( depending on the glycerol weight fraction and temperature), densities from $1.05$ g/ml to $1.25$ g/ml, and an average surface tension of $65$ mN/m (as measured with a pendant drop set-up).

\subsection {Methods}
\subsubsection {Drop impact experiments}

\begin{figure}
\includegraphics[width=8.5cm]{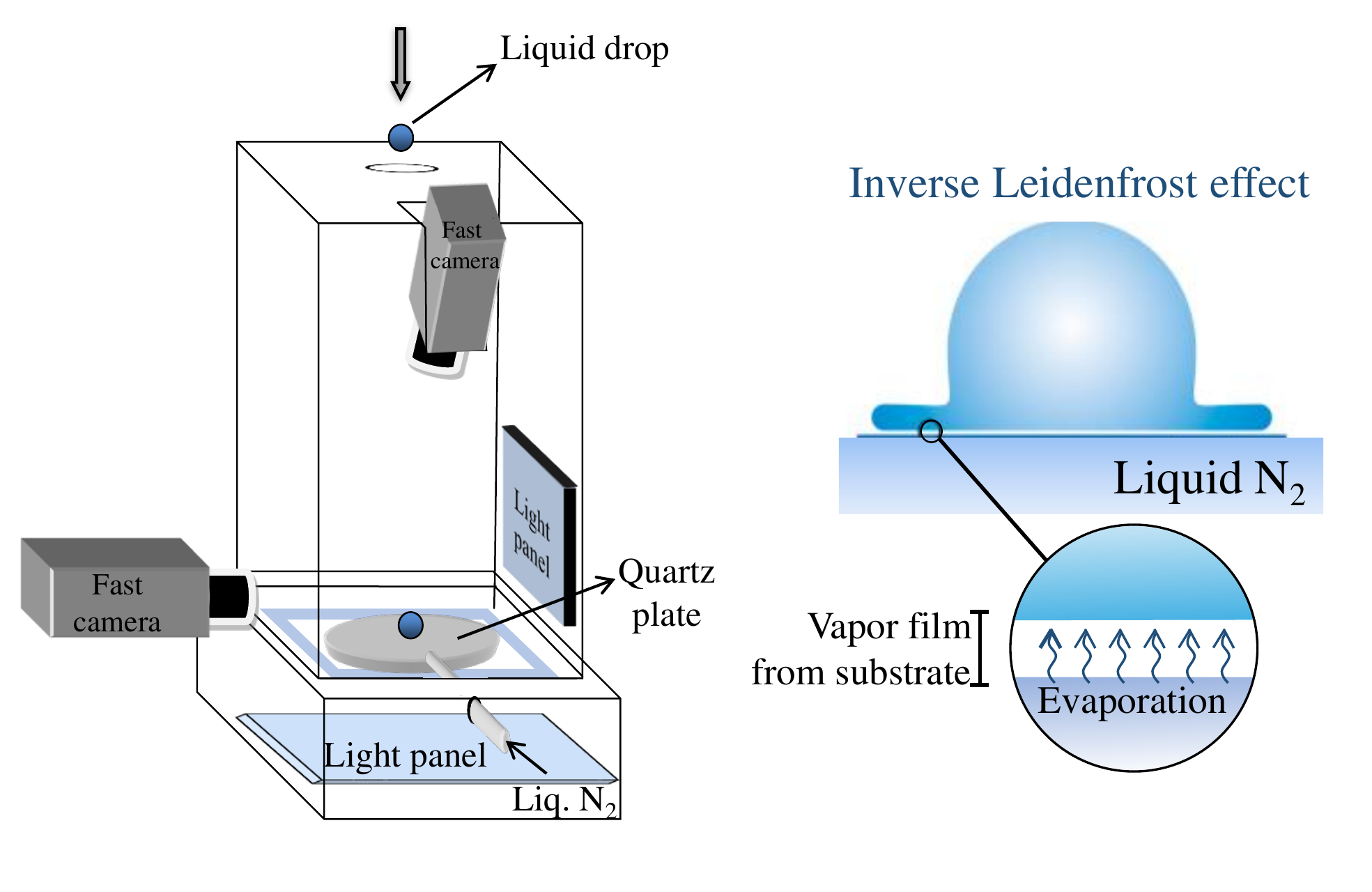}
\caption{Left: Schematic illustration of the impact experiments setup showing a drop falling on a liquid nitrogen thin layer. The expansion event is recorded using two fast cameras allowing concomitant top and side visualizations. Right: adapted from \cite{Antonini2013}, schematic of the inverse Leidenfrost effect at the origin of the shear free expansion.}
\label{figure:1}
\end{figure}

To substantially eliminate the role of friction or adhesion with the solid surface on the impact dynamics, we work under inverse Leidenfrost conditions. This is achieved by impacting a drop at ambient temperature $T$ (between 18.5 and $22.5^ {\circ}$C)  on a polished quartz slide covered with a thin layer of liquid nitrogen ($\rm{N_2}$) at $ T=-196 ^\circ $C (see fig.~\ref{figure:1}). The setup is described elsewhere~\cite{Arora:2018ei}. Upon impact of the drop, a vapor cushion forms at the liquid interface due to the evaporation of $\rm{N_2}$, providing a unique scenario of non-wetting slip conditions that eliminates shear viscous dissipation~\cite{Antonini2013, Chen:2016fo}.  The rare cases where the impacting drop comes in direct contact with the surface, with instantaneous freezing are eliminated, so  the nitrogen vapour film keeps the liquid drop separated from the surface for  all the reported experiments.  Before each impact, the quartz slide is first cleaned by blowing $\rm{N_2}$ gas and then a thin layer (typical thickness $50$ nm as measured by ellipsometry) of liquid $\rm{N_2}$ is deposited on the slide. The liquid is  injected from a syringe pump with a flow rate of $1$ ml/min through a needle placed above the target, from the side as shown in Fig.1. The diameter of the falling drop is constant, $d_0=3.9\pm0.2$ mm, as measured by image analysis and confirmed from the drop mass. The drop falls from a height $ h=91$ cm, yielding an impact velocity $v_0=\sqrt{2gh}=4.2\: \mathrm{ms^{-1}}$($g $ is the acceleration of gravity).The drop impact is recorded from the top (Fig.2a) using a high-speed camera (Phantom V 7:3) operated at $6700 \, \mathrm{frames/s}$ with a resolution of $800 \times 600\:  \mathrm{pixels^2}$. The angle between the camera axis and the horizontal plane is fixed to about $10^\circ $. A second high-speed camera (Phantom miro M310), operated at $3200 \: \mathrm{frames/s}$ with a resolution of $1280 \times 800\:  \mathrm{pixels^2}$, is eventually used simultaneously to record a side view (Fig.2b).

\subsubsection {Image analysis}

\begin{figure}
\includegraphics[width=8.5cm]{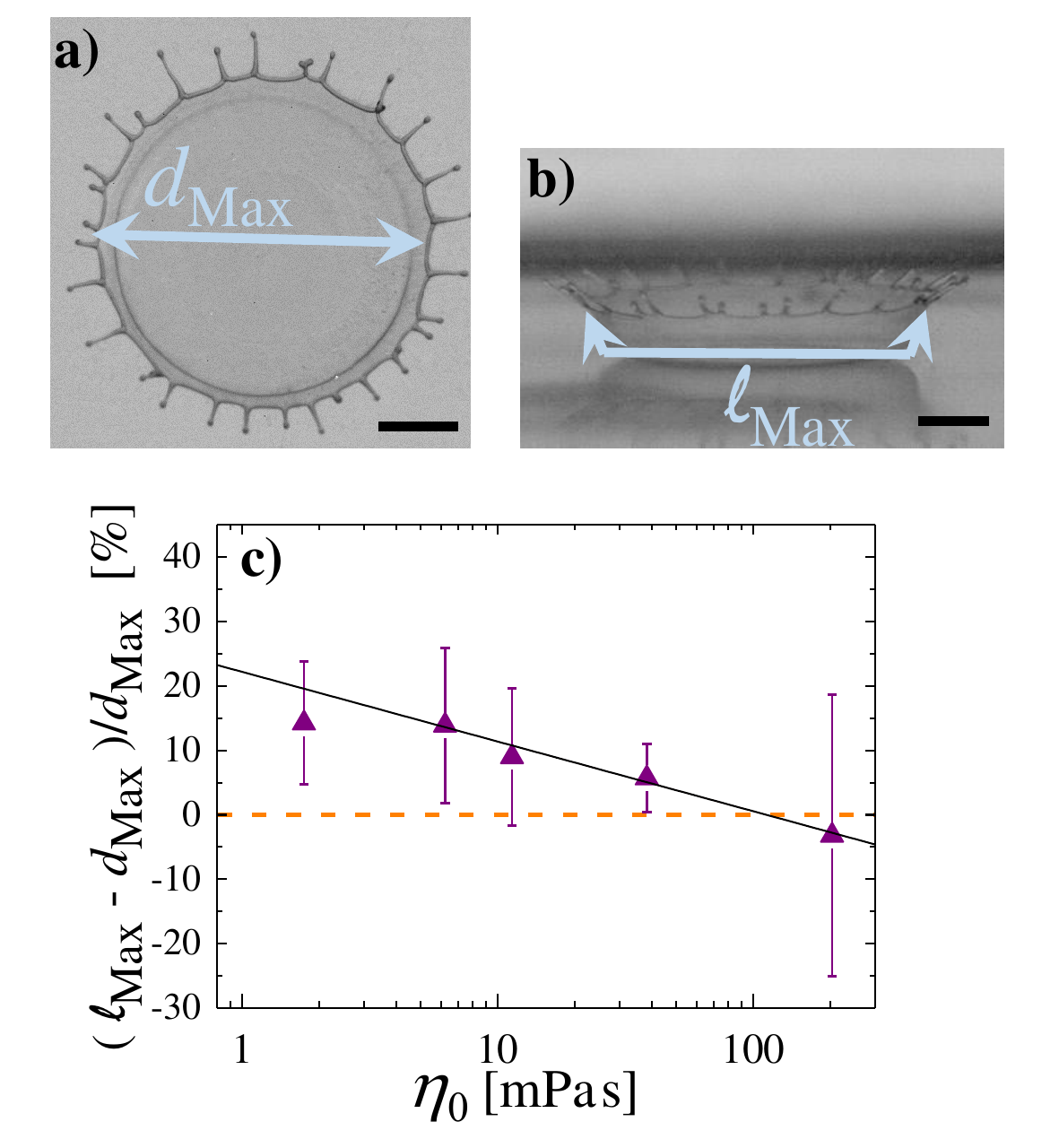}
\caption{(a): Top view and (b): side view snapshots of the maximum expansion of a low viscous Newtonian sample (22 $\rm{wt}\%$ Glycerol/water mixture), revealing that the effective diameter $d_{\rm{Max}}$ from the top view underestimates the maximum expansion of the sheet $\ell_{\rm{Max}}$.(c): Relative side view correction effect (in \%) as a function of  shear viscosity of Newtonian glycerol/water. The line is an empirical fit of the data points (symbols).}
\label{figure:2}
\end{figure}

The time evolution of the sheet size is measured with ImageJ software by analyzing top view images. We first subtract the background image from the expansion movie and highlight the rim by a binary thresholding. This allows us to determine the sheets contour and measure its area $A$.  Note that $A$ does not include   the fingers emanating from the rim of the expanding
film;   they may appear   for low viscous  samples (Fig.\ref{figure:2}), but do not develop for more viscous ones ( Fig.\ref{figure:4}). An apparent sheet diameter is simply deduced: $d=\sqrt{\frac{4A}{\pi}}$. The results are  obtained by averaging for each sample the time evolution of the sheet diameter from three different experiments. Note however that corrections have to be performed for low viscosity samples. Indeed, side view images reveal the occurrence of a  so-called corona splash (Fig.\ref{figure:2}b) \cite{Josserand:2016jf} , for low viscous Newtonian samples. This implies that the routine standard analysis using top view images would underestimate the maximum expansion diameter. The side view allows one to evaluate the actual diameter of the sheet (see $\ell_{\rm{Max}}$  in Fig.~\ref{figure:2}b). The fractional underestimation, defined as $\frac{\ell_{\rm{Max}}-d_{\rm{Max}}}{d_{\rm{Max}}}$, with $d_{\rm{Max}}$ the maximum diameter measured from top-view images, is plotted as a function of the samples zero-shear viscosity, $\eta_0$, in Fig.\ref{figure:2}c. We find that the fractional underestimation decreases logarithmically with $\eta_0$, from about $22 \rm{wt}\%$ for the lowest viscosity sample and vanishes for $\eta_0 \geq 100$ mPa.s. Hence, in the following, quantitative corrections are made for the maximum expansion diameter based on the empirical logarithmic law. For samples with a shear viscosity larger than $100$ mPa, the sheet remains flat and no correction is needed.

\subsubsection {Rheology}

Linear viscoelastic and steady shear viscosity measurements are performed with a MCR302 stress-controlled rheometer (Anton Paar, Austria), operating in the strain-control mode and equipped with a stainless steel cone and plate geometry with a diameter of $50$ mm, cone angle of $\ 1^\circ$ and truncation of $101$ $\mu$m. Temperature control ($\pm 0.2^\circ$ C) is achieved by means of a Peltier element.
The linear viscoelastic spectra are obtained by applying a small amplitude sinusoidal strain, such that data are obtained in the linear regime ($\gamma = 10\%$), varying the angular frequency, $\omega$, from $100$ to $0.01$ rad/s, and measuring the storage, $G'(\omega)$, and loss, $G"(\omega)$, moduli. The complex viscosity is calculated from the linear viscoelastic spectra as $ | \eta^\star (\omega)|=\frac{\sqrt{(G'^2 (\omega)+G"^2 (\omega) )}}{\omega} $. The steady shear viscosity, $\eta({\dot{\gamma}})$  is  measured by applying a ramp of steady shear rate varying from $0.01$ to $1000\, \mathrm{s^{-1}}$.

\section {Results}
\subsection{Shear Rheology} \label{shrheo}

\begin{figure}
\includegraphics[width=8.5cm]{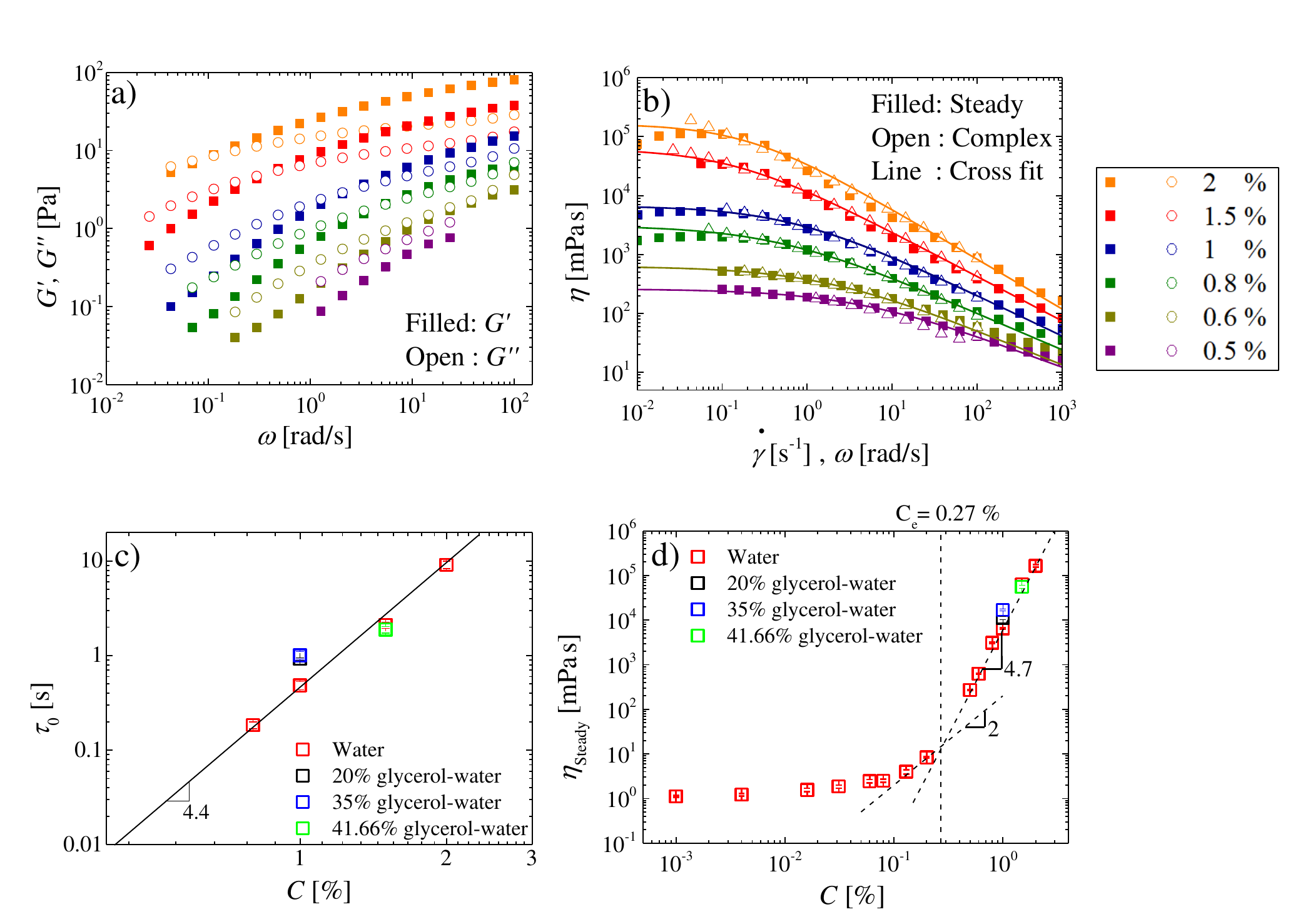}
\caption{(a) Frequency dependence of the storage ($G'$) and loss ($G''$) moduli, and (b) Complex viscosity as a function of frequency (open symbols) and steady shear viscosity as a function of shear rate (filled samples), and fits (lines) using the Cross equation (Eq.1), for samples with different PEO concentrations $C$ as indicated in the legend. (c)  Evolution with $C$ of the terminal relaxation time, and (d)  of  the zero shear viscosity.}
\label{figure:3}
\end{figure}

Figure~\ref{figure:3}a shows the dynamic moduli as a function of oscillatory frequency for aqueous polymer solutions with various concentrations. For $C>0.6 \rm{\rm{wt}}\%$, the crossover of $G'$ and  $G"$ marks a characteristic relaxation time $\tau_0$, which is the best estimate for the onset of the terminal regime. Figure~\ref{figure:3}c shows that $\tau_0\sim C^{0.44}\,\mathrm{s}$. This scaling exponent is in the range that have been reported for high molecular weight PEO aqueous solutions \cite{Ortiz:1994kh}. Results for samples prepared with mixtures of water and glycerol are consistent with those obtained for pure water samples (fig.~\ref{figure:3}c). 

The zero shear viscosity, $\eta_0$, varies by more than $5$ orders of magnitude from $1$ mPa.s to $10^5$  mPa.s for the samples investigated. The variation of $\eta_0$ with polymer concentration, $C$, reveals the two expected regimes (fig.~\ref{figure:3}d): an unentangled regime for $C<0.27\,\%$, where the viscosity increases slowly with the polymer concentration, and an entangled regime at larger concentration, where $\eta_0\propto C^{4.7}$, in agreement with predictions by scaling arguments based on the tube model \cite{Rubinstein:2003bk}.

Figure~\ref{figure:3}b depicts the complex viscosity, $| \eta^\star (\omega)|$ as a function of frequency, along with the steady shear viscosity, $\eta(\dot{\gamma})$ as a function of shear rate, $\dot{\gamma}$. The nice collapse of the dynamic and steady data validates the Cox-Merz rule~\cite{Cox:1958jn}. We find that all samples are strongly shear-thinning and that an empirical fit by means of the Cross model provides a good description of the shear-thinning behavior of PEO solutions (continuous lines in fig.~\ref{figure:3}b)~\cite{Cross:1965bm}.

\begin{equation}
\label{eqn:Crossfit}
\eta_s(\dot{\gamma})=\eta_\infty+ \frac{\eta_0-\eta_\infty}{1+(k\dot{\gamma})^n}
\end{equation}

Here, $\eta_\infty$ is the viscosity at very large shear rate that we set equal to the solvent viscosity  (water or water/glycerol mixture), $\eta_0$ is the zero-shear viscosity (plotted in fig.~\ref{figure:3}d), $n$ is the shear-thinning exponent and the parameter $k$ is the inverse of a critical shear rate that delimitates a Newtonian regime from a shear-thinning regime. We find that the shear-thinning exponent $n$ increases with increasing $C$ from $0.59$ to $0.85$. Moreover, for all concentrations the fitting parameter $k$ is monotonically increasing with $C$, similarly to the characteristic relaxation time $\tau_0$, and it marks the onset of shear-thinning.

\subsection{Drop impact experiments}

\begin{figure}
\includegraphics[width=8.5cm]{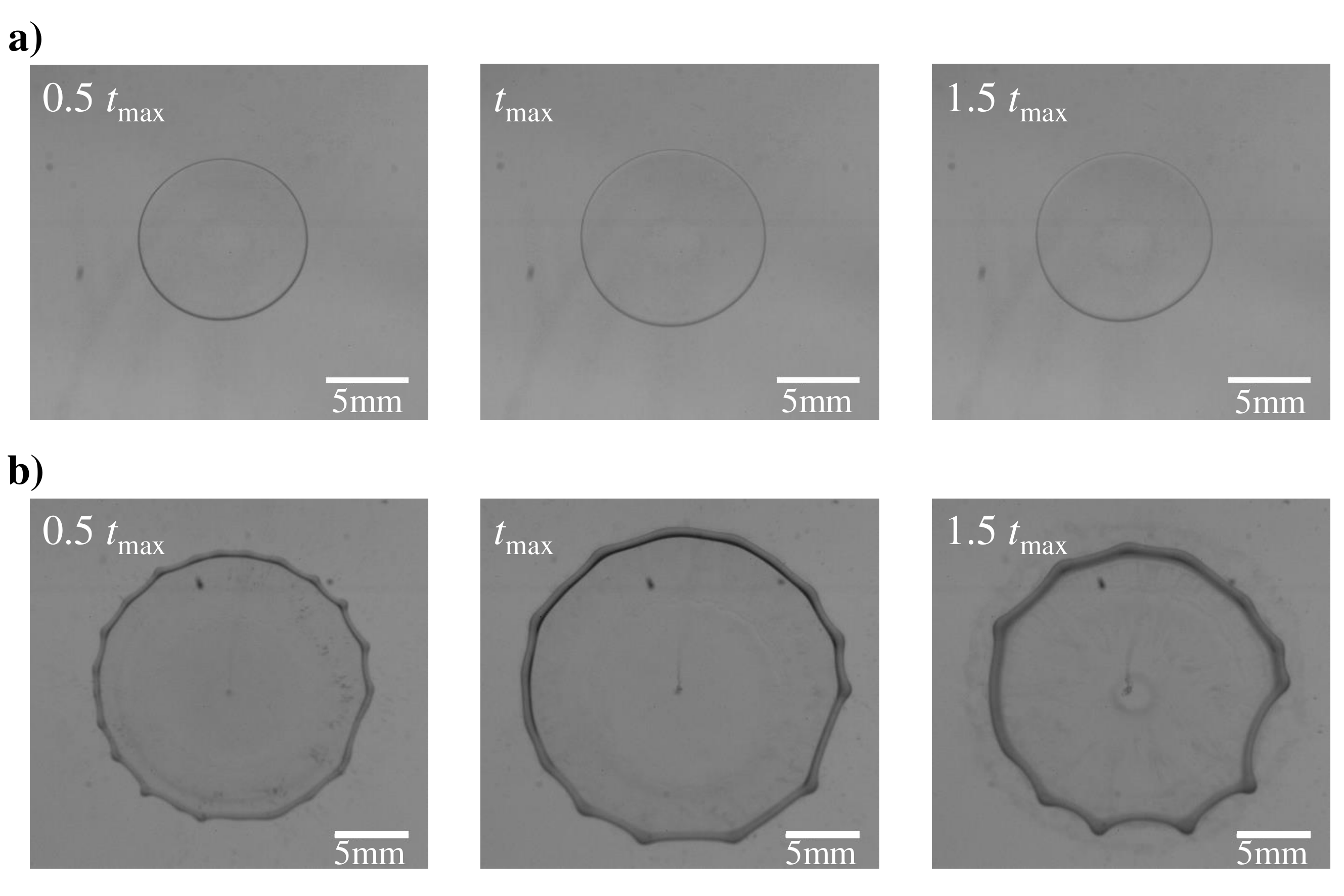}
\caption{Snapshots taken at different times, as indicated, during the expansion and retraction of the sheet for (a) a Newtonian silicon oil with  shear viscosity $\eta_0=658$ mPa.s ;  maximum expansion reached at time $t_{max}=5.22 $ ms and (b) a PEO solution with $C=0.6 \rm{wt}\%$ and $\eta_0=628$ mPa.; maximum expansion reached at time $t_{max}=5.02 $ ms. The bar sets the scale.}
\label{figure:4}
\end{figure}

\begin{figure}
\includegraphics[width=8.5cm]{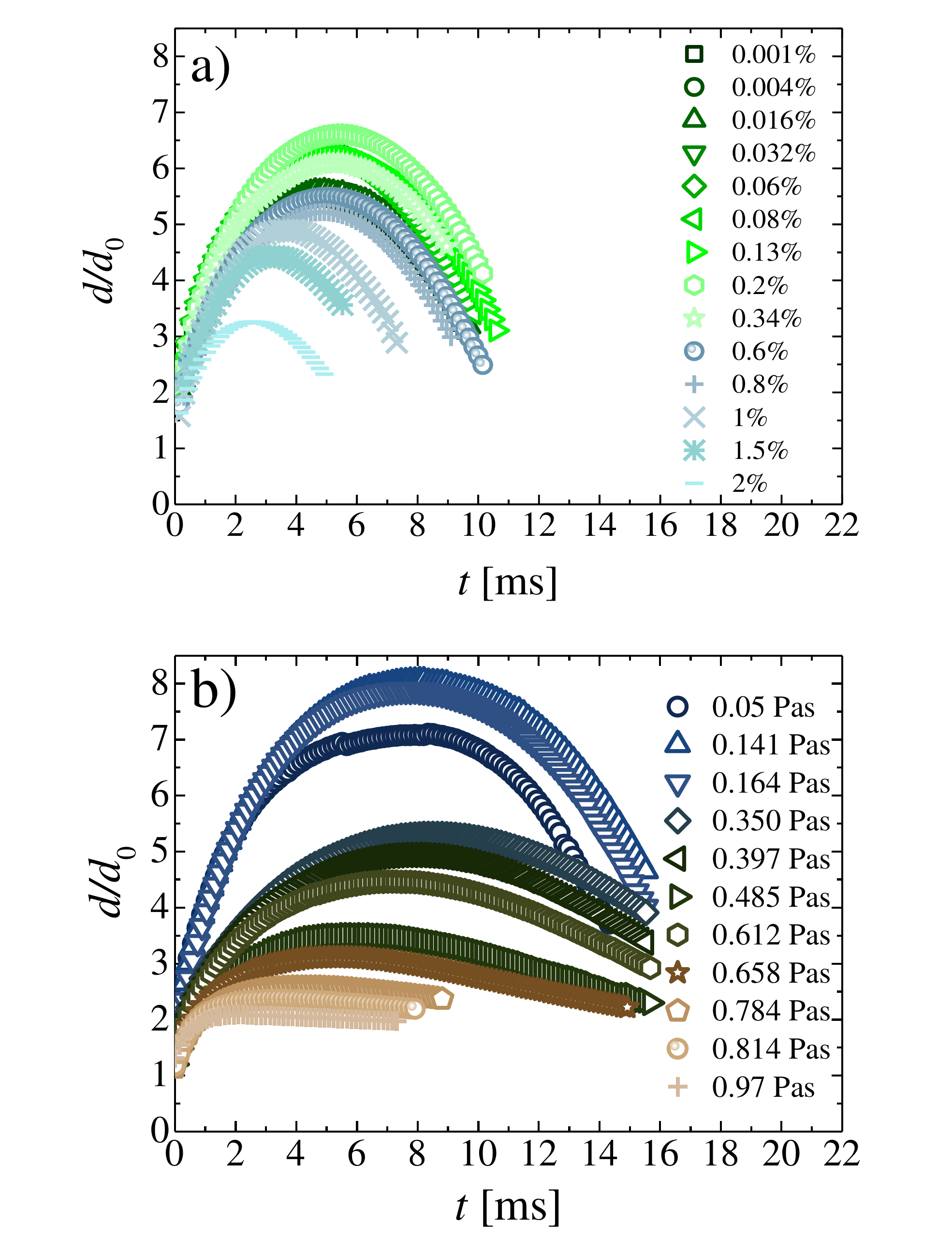}
\caption{Time evolution of the sheet diameter normalized by the initial drop diameter for (a) PEO solutions with different concentrations and (b) silicon oils with different shear viscosities, as indicated in the legends. The origin of the time is taken at the drop impact.}
\label{figure:5}
\end{figure}

Once hitting the repellent surface, the drop  expands freely in air until reaching a maximum expansion. It then retracts because of surface tension. This corresponds  to an   an  axisymmetric biaxial extensional  flow . The overall behavior is illustrated in figure~\ref{figure:4} that depicts snapshots of the drop after its impact for a PEO solution with $C=0.6 \%$  ($\eta_{0} = 628$ mPa.s) and for a Newtonian sample of comparable zero-shear viscosity ($\eta_0=658$ mPa.s). The two samples display strikingly different behavior: the viscoelastic fluid drop expands much more than the Newtonian drop and moreover forms a thicker rim. More quantitatively, we show (fig.~\ref{figure:5}a,b) selective raw data for the time evolution of the effective sheet diameter normalized by the original drop, $\frac{d}{d_{0}}$, for PEO solutions at different concentrations and Newtonian liquids  (here silicon oils, but water/glycerol mixtures exhibit the same behavior) with different viscosities. The origin of time is chosen at the time when the drop hits the liquid nitrogen layer. Expansion and retraction regimes are shown, yielding a bell shape for the curves. We note that the curves for the viscoelastic fluids are very symmetric as opposed to the ones for the more viscous Newtonian fluids, where a very long retraction regime is measured. These findings deserve deeper investigations in the future, but in the following, we focus on the maximum expansion diameter $d_{\rm{Max}}$.

For a biaxial extensional flow,during the expansion of free liquid sheets after impact on a repellent surface, the relevant viscosity is the biaxial extensional viscosity  defined as $\eta_{\rm{B}}=\frac{\sigma_{rr}-\sigma_{zz}}{\dot{\epsilon}}$, where $\dot{\epsilon}$ is the strain rate, and $\sigma_{rr}$  and $\sigma_{zz}$ are the stress tensor components in cylindrical coordinates. 

\begin{figure}
\includegraphics[width=8.5cm]{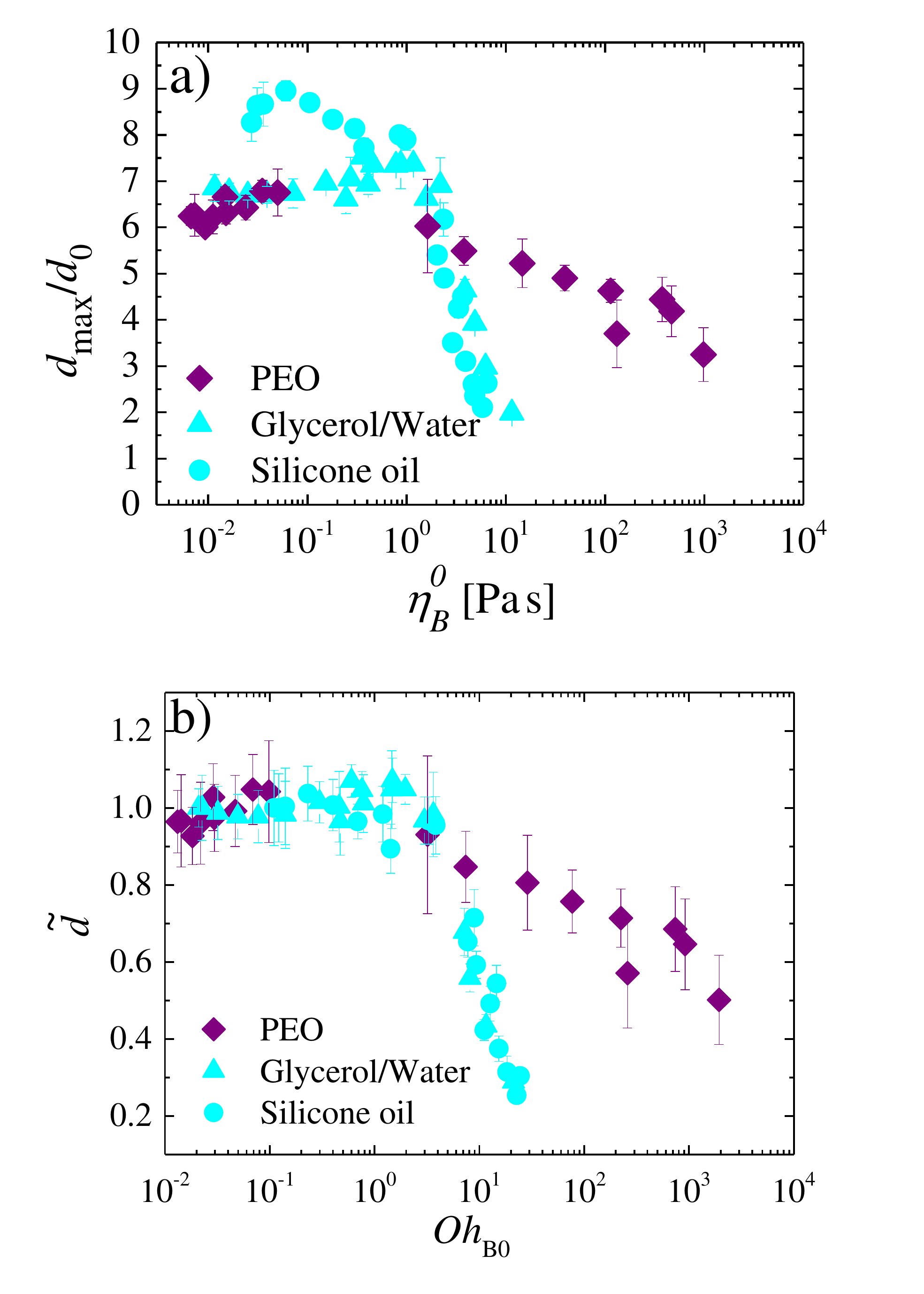}
\caption{ a) Maximum diameter of the sheet normalized by the initial drop diameter as a function of the biaxial extensional viscosity, and b) Normalized maximum expansion factor as a function of the biaxial extensional Ohnesorge number, for PEO solutions and for Newtonian liquids.}
\label{figure:6}
\end{figure}

For a Newtonian fluid the constant biaxial extensional  viscosity  $\eta_{\rm{B}}=\eta_{\rm{B}}^0=6\eta_0$, where $\eta_0$ is the zero-shear viscosity and $6$ is the Trouton ratio~\cite{Macosko:1994bk}. As a first order analysis aiming at rationalizing the expansion dynamics of sheets of Newtonian and thinning fluids, we consider the maximum expansion factor, $d_{\rm{Max}}/d_0,$ with $d_0 $ the initial drop diameter and plot (fig.~\ref{figure:6}a) this quantity as a function of the biaxial extensional viscosity $\eta_{\rm{B}}^0$. Note that  the shear viscosity is measured  at the room temperature of the impact experiment, this is  of particular importance for water/glycerol mixtures which exhibit a strong temperature dependant viscosity \cite{Oberstar:1951im}.
Similarly to findings for Newtonian fluids impacting a small solid target~\cite{Arora:2016bu}, two regimes are observed for both polymer solutions and Newtonian fluids. At low $\eta_{\rm{B}}^0$ a capillary regime prevails, where the maximum expansion is mainly driven by a balance between surface tension forces and inertial  forces, with viscous dissipation being negligible. Hence, this regime is characterized by a plateau. By contrast, in the viscous regime at higher $\eta_{\rm{B}}^0$, $d_{\rm{Max}}$ decreases monotonically with increasing viscosity.

To get further insight into the observed behavior, we use the normalized maximum expansion factor, $\tilde{d}$ adopting the same definition as in Refs.~\cite{An:2012gc, Arora:2016bu}:

\begin{equation}
\label{eqn:dtilde}
\tilde{d}=\frac{d_{\rm{Max}}}{d^{\rm{cap}}_{\rm{Max}}}
\end{equation}

where $d_{\rm{Max}}^{\rm{cap}}$ is the maximum expansion diameter in the capillary regime (at low viscosity). This normalized quantity allows us to compare drops with different initial sizes and different surface tensions. In addition, to account for different surface tensions for different samples, data are plotted against the biaxial extensional Ohnesorge number, $Oh_{\rm{B0}}$, the ratio of biaxial extensional viscous forces to inertial and  surface tension forces:

\begin{equation}
\label{eqn:Ohnesorge}
Oh_{\rm{B0}}=\frac{\eta_{\rm{B}}^0}{\sqrt{\rho\gamma d_0}}
\end{equation}

 with $\rho$ the sample density and $\gamma$ the surface tension. Figure~\ref{figure:6}b shows the evolution of $\tilde{d}$  with $Oh_{\rm{B0}}$ for both polymer solutions and Ne\rm{wt}onian fluids. The data for the two types of Newtonian samples overlap nicely onto a master curve and exhibit a capillary regime  (for $Oh_{\rm{B0}}\lesssim Oh_{\rm{B0}}^c=2$) characterized by a plateau, followed by a biaxial extensional viscous dissipation regime. We find that for the thinning fluids, the  onset of the viscous regime takes place at approximatively the same critical biaxial extensional Ohnesorge number $Oh_{\rm{B0}}^c$  as for Newtonian liquids. Interestingly, however, $\tilde{d}$ decreases much more gradually with $Oh_{\rm{B0}}$ as compared to Newtonian fluids in the viscous regime. This clearly suggests the importance of a biaxial extensional thinning of the polymer solutions in the viscous dissipation regime. In the next section, we provide a rationalization for the biaxial extensional expansion dynamics of sheets produced with Newtonian and non-Newtonian fluids.

 \section {Discussion}
 \subsection {Rationalization of biaxial extensional thinning}

\begin{figure}
\includegraphics[width=8.5cm]{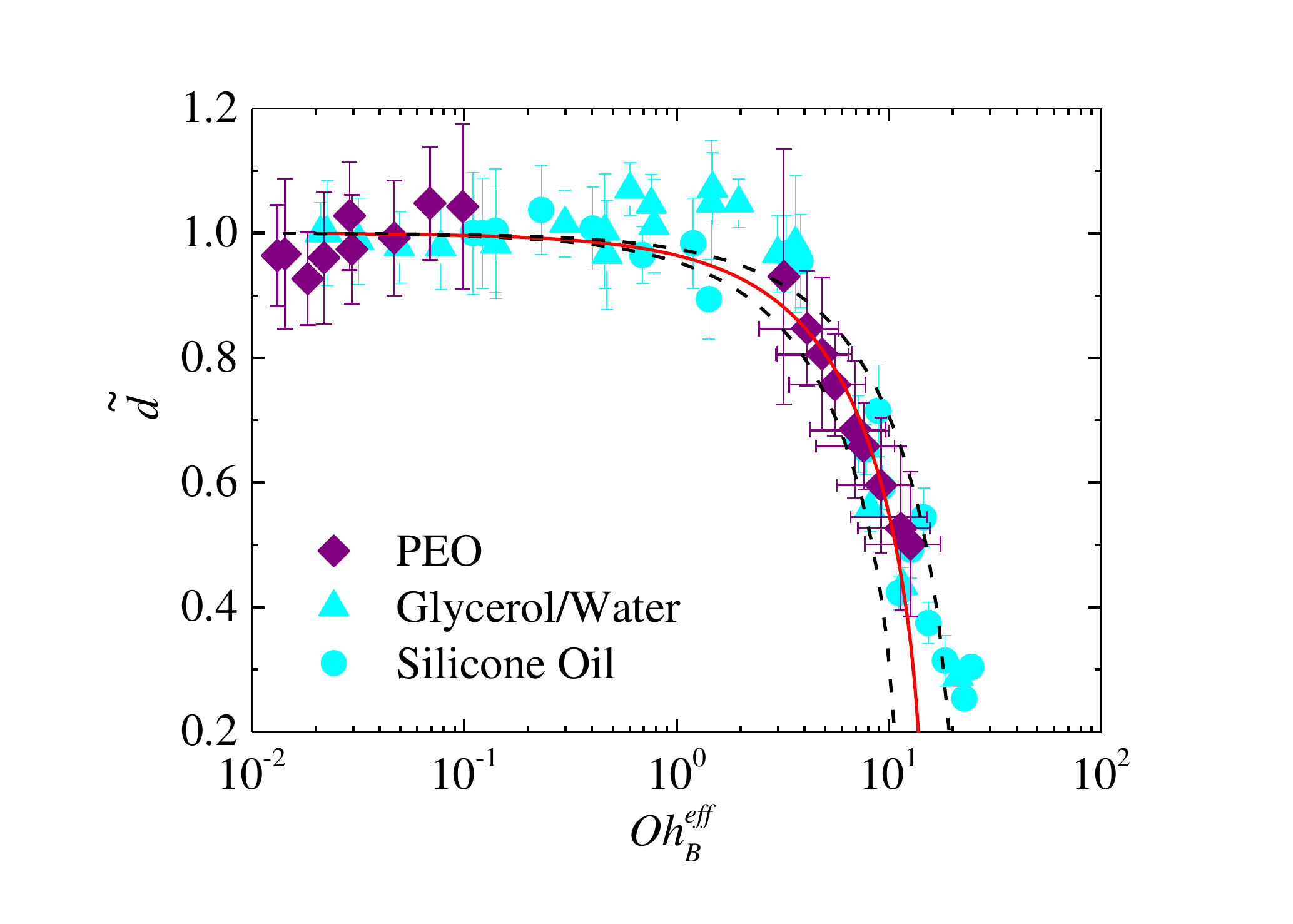}
\caption{Normalized maximum expansion factor as a function of the effective  biaxial extensional Ohnesorge number for  polymer solutions and for the two classes of Newtonian samples. The thin continuous line is the best fit with Eq. 11, and the dashed lines are used to evaluate error bars on the fit parameter.}
\label{figure:7}
\end{figure}

From the data of figure~\ref{figure:6}b, one can easily define an effective biaxial extensional thinning viscosity $\eta_{\rm{B}}^{\rm{shift}}$ for polymer samples belonging to the viscous regime ($Oh_{\rm{B0}}> Oh_{\rm{B0}}^c$) by shifting horizontally the experimental data point so that they fall on the master curve found for Newtonian samples. The shifted values are discussed in Fig.9 below. No shift is performed in the capillary regime ($Oh_{\rm{B0}} < Oh_{\rm{B0}}^c$) since viscous dissipation is not relevant. Doing so we build a master curve (fig.~\ref{figure:7}) for the maximum expansion $\tilde{d}$ as a function of $Oh_{\rm{B}}^{\rm{Thin}}$ for all types of samples, where  the effective biaxial extensional Ohnesorge number $Oh_{\rm{B}}^{\rm{Thin}}= Oh_{\rm{B0}}$ for Newtonian samples and viscoelastic thinning samples in the capillary regime and   $Oh_{\rm{B}}^{\rm{Thin}}= \frac{\eta_{\rm{B}}^{\rm{shift}}}{\eta_{\rm{B}}^0}{ Oh_{\rm{B0}}}$ for visoelastic thinning samples in the viscous regime. Below we rationalize the shifting of Fig.7 and the use of biaxial extensional viscosity of non-Newtonian fluids.

\subsubsection{Determination of the pertinent strain rate} \label{strainrate}

Experimentally, measuring properly the biaxial extensional viscosity is a challenging task, especially for relatively low viscosity fluids such as the present polymers~\cite{Maerker:1974jx, Joye:1972pd,Venerus:2010jl, Johnson2016:bn, Huang:1993lm, Hachmann:2003kl,Cathey:1988ip}. In order to rationalize  $\eta_{\rm{B}}^{\rm{shift}}$, the first step is to properly account for the deformation rate experienced by sheets during their expansion in air. Here, we provide an estimate.

\begin{figure}
\includegraphics[width=8.5cm]{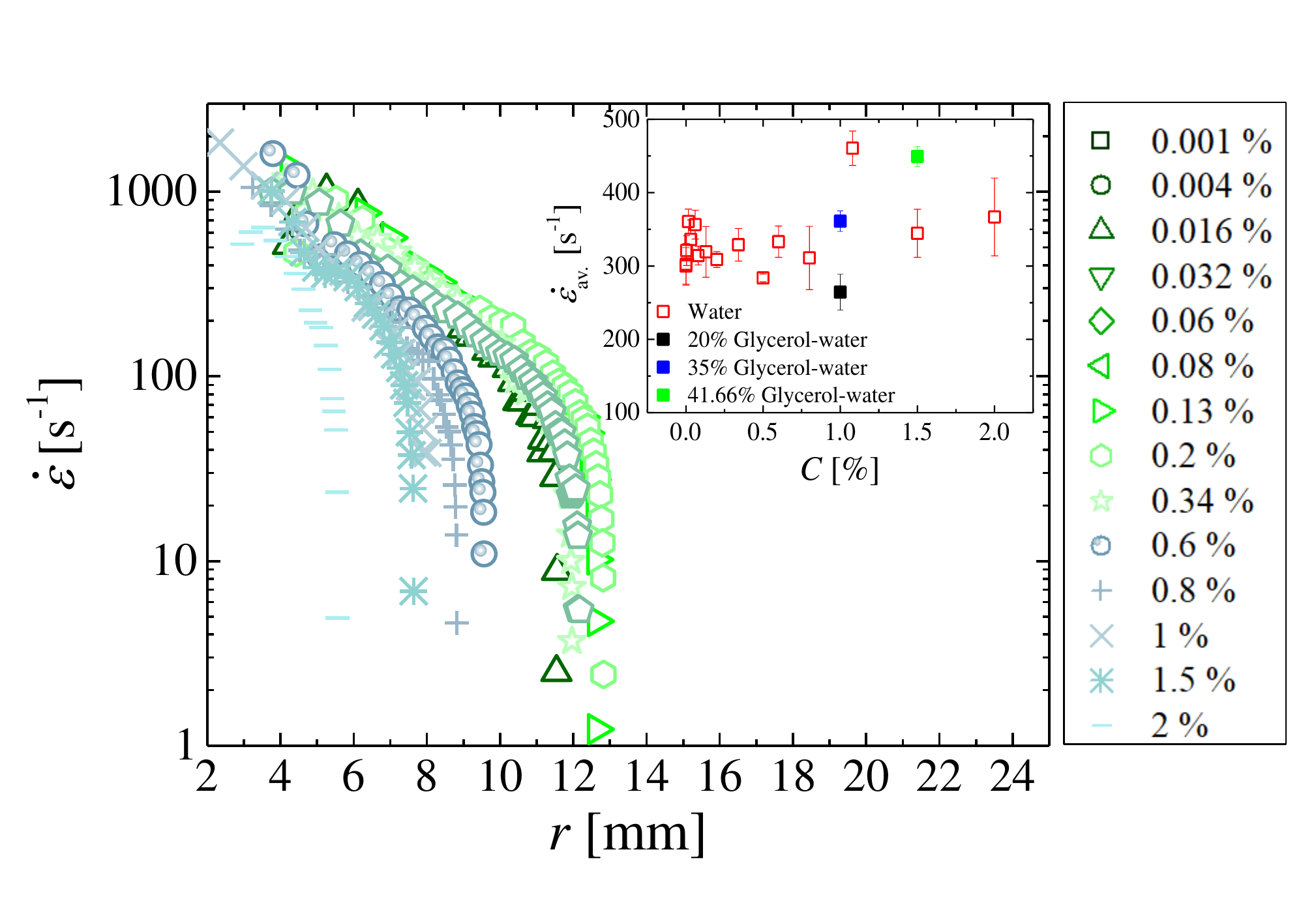}
\caption{Evolution of the expansion strain rate as a function of the sheet radius during its expansion for aqueous PEO solutions at different concentrations, as indicated in the legend.Inset: Average strain rate calculated according to Eq. 4 as a function of concentration. Error bars represent the standard deviation from three different experiments.}
\label{figure:8}
\end{figure}

The  effective strain rate, defined as $\dot{\varepsilon}=\frac{1}{d}\frac{\partial d}{\partial t}$, is not constant in the expansion regime, but decreases with time and vanishes at maximum expansion (fig.~\ref{figure:8}). Note that the film expansion is a time-dependent problem but here we focus on the maximum diameter (end of expansion process) and measure  the average rate  experienced   by the sheet during the expansion .~The average value for the strain rate  in the expansion regime is then calculated as:

\begin{equation}
\label{eqn:strainrate}
\dot{\varepsilon}_{\rm{av}}=\frac{\int_0^{r_{\rm{Max}}}r\dot{\varepsilon}dr}{\int_0^{r_{\rm{Max}}}rdr}
\end{equation}

Here $r_{\rm{Max}}=\frac{d_{\rm{Max}}}{2}$ is te radius of the sheet at its maximum expansion.We get an average strain rate for each solutions that it is used for the rest of the analysis. Note, however, that within experimental errors, ($\simeq 15\%$ as shown in the inset of Fig.~\ref{figure:8}), the effective strain rate does not vary a lot with concentration. Indeed, we obtain a value of $338 \pm 49\,\mathrm{s^{-1}}$ by averaging over all samples, including the PEO solutions prepared in glycerol/water mixtures.

\subsubsection{biaxial extensional viscosity}

Given the difficulty in obtaining reliable experimental data for the biaxial extensional viscosity $\eta_{\rm{B}}(\dot{\varepsilon})$ with our samples, we attempt at  providing reasonable estimations To this end, we rely on two pioneering experimental works for the measurements of the biaxial extensional viscosity of viscoelastic solutions~\cite{Walker1996:kk, Venerus:2019cj}, where similar scaling have been found in spite of using different techniques and different samples, i.e. wormlike micelles in Ref.~\cite{Walker1996:kk} and concentrated  polymer solutions in Ref.~\cite{Venerus:2010jl}. At low Weissenberg numbers ($Wi=\dot{\varepsilon} \tau_0 < 1 $), i.e., for rates lower than the inverse of the terminal relaxation time, $\tau_0$, the biaxial extensional viscosity is independent of the rate and follows the expectation for a Newtonian fluid: $\eta_{\rm{B}}=\eta_{\rm{B}}^0=6\eta_0$ with $\eta_0$ the zero-shear viscosity. By contrast, when $Wi>1$, the biaxial extensional viscosity decreases with rate, $\eta_B\sim\dot{\varepsilon}^{-p}$ with a thinning exponent $p=0.5$~\cite{Venerus:2010jl}. This scaling has been also predicted by Marrucci and Ianniruberto~\cite{Marrucci:2004cd} using a tube-based model for polymer melts hence,  pointing out the universality of the biaxial extensional thinning behavior.

\begin{figure}
\includegraphics[width=8.5cm]{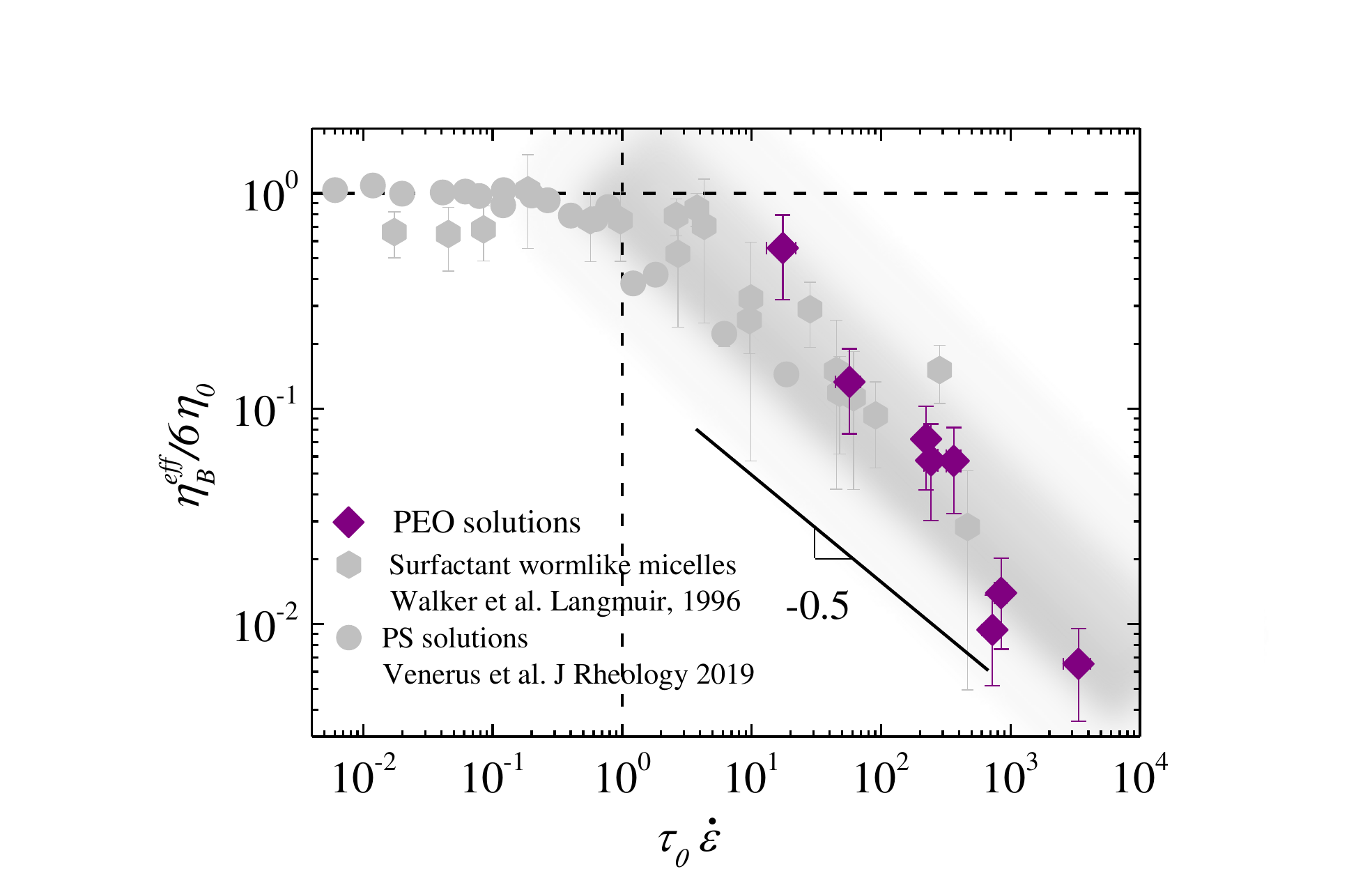}
\caption{Effective biaxial extensional viscosity of PEO solutions, extracted from the manual shift of the data in Fig.7, normalized by the plateau biaxial extensional viscosity as a function of the Weissenberg number, plotted together with literature data as indicated in the legend. A unique master biaxial extensional flow curve is obtained with a thinning exponent of  $-0.5$ in agreement with theoretical predictions (see text). The grey zone highlights  the -0.5 slope.}
\label{figure:9}
\end{figure}

Based on the similar linear viscoelastic response for the PEO solutions and those investigated in~\cite{Walker1996:kk, Venerus:2019cj}, we expect our samples to exhibit a similar behavior for the biaxial extensional viscosity as a function of expansion rate. Thus, for each impact experiment in the viscous regime, we define an effective Weissenberg number as: $Wi^{\rm{eff}}=\tau_0 \dot{\varepsilon}_{\rm{av}}$, where $\tau_0$ is given by linear shear rheology measurements for the data showing crossover between $G^{'}$ and $ G^{''} $ and extrapolated, according to the power law presented in (fig.~\ref{figure:3}c), for data without the crossover ({$0.5 \rm{wt}\%$} and {$0.6 \rm{wt}\%$} ); $\dot{\varepsilon}_{\rm{av}}$ is measured from experiments (fig.~\ref{figure:8}b). The effective biaxial extensional thinning  is  characterized  by the experimental data points ($Wi^{\rm{eff}}, \eta_{\rm{B}}^{\rm{shift}}$), where $\eta_{\rm{B}}^{\rm{shift}}$ are obtained from fig.~\ref{figure:7}.
We show in figure~\ref{figure:9} the  variation of the normalized effective biaxial extensional viscosity of PEO  solutions $\eta_{\rm{B}}^{\rm{shift}}/(6\eta_0)$  as a function of the effective Weissenberg number $Wi^{\rm{eff}}$ obtained from drop impact experiments in the viscous regime. On the same plot, we report the experimental data of Refs.~\cite{Walker1996:kk, Venerus:2019cj}. We find a remarkable agreement between our shifted values  and those from the literature, even though the expansion sheet dynamics that result from the impact of drops are non-stationary, supporting our simple approach.

\subsection{Rationalization of the maximal expansion by considering the biaxial deformation}

To go one step further, we provide below scaling laws to account semi-quantitatively for the dependence of the maximum expansion on the biaxial extensional viscosity (fig.\ref{figure:7}). We restrict our analysis to Newtonian samples. Indeed, for the normalized maximum expansion factor,   each non Newtonian sample in  the viscous regime, can be replaced by a Newtonian sample exhibiting the same normalized maximum expansion factor a shown before.


 We adopt an energy conservation balance and first consider the capillary regime, for which the initial kinetic energy is assumed to be fully converted into surface energy at the maximum expansion of the sheet:


\begin{equation}
\label{eqn:energy1}
\frac{1}{2}mv_0^2\simeq2\pi \gamma (d^{\rm{cap}}_{\rm{Max}})^2
\end{equation}
where $m=\rho \pi d_0^3/6$ is the mass of the drop and $d^{\rm{cap}}_{\rm{Max}}$ is the diameter at maximum expansion of the sheet in the capillary regime where viscous dissipation is negligible. In the viscous regime, we need to add to the right hand side of Eq.\ref{eqn:energy1} a term accounting for the viscous dissipation, which here is assumed to result only from the biaxial extensional deformation:

\begin{equation}
\label{eqn:energy2}
\frac{1}{2}mv_0^2\simeq2\pi \gamma d^2_{\rm{Max}} +E_{\rm{B}}
\end{equation}

Combining Eqs.\ref{eqn:energy1} and~\ref{eqn:energy2}, we obtain for the normalized maximum expansion factor, $\tilde{d}$

\begin{equation}
\label{eqn:dtilde}
\tilde{d}=\frac{d_{Max}}{d^{\rm{cap}}_{\rm{Max}}}=\sqrt{1-\frac{2E_{\rm{B}}}{mv_0^2}}
\end{equation}

with $E_{\rm{B}}$ the biaxial extensional extensional energy dissipated during the process of sheet expansion. To a first approximation, $E_{\rm{B}}$ can be written as:

\begin{equation}
\label{eqn:dissip1}
E_B\approx\ \int_0^{t_{\rm{Max}}}dt  \int_V \sigma_{\rm{B}}(\dot{\varepsilon})\dot{\varepsilon} dV
\end{equation}

where $V=\frac{\pi d_0^3}{6}$ is the volume of the drop, $\sigma_{\rm{B}}(\dot{\varepsilon})= \eta_{\rm{B}} \dot{\varepsilon}$, and $t_{\rm{Max}}$ is the time to reach maximum expansion.~Hence, assuming also a volume conservation with a uniform  thickness sheet , Eq.~\ref{eqn:dissip1} can be rewritten as:

\begin{equation}
\label{eqn:dissip2}
E_{\rm{B}}\approx \eta_{\rm{B}}\frac{\pi d_0^3}{6} \int_0^{t_{\rm{Max}}} \biggl ( \frac{1}{d}\frac{\partial d}{\partial t}\biggr ) ^2 dt
\end{equation}
Using a simple first order scaling approach to calculate the strain rate, we consider $\frac {\partial d}{\partial t}\approx \frac{d_{\rm{Max}} -d_0}{t_{\rm{Max}}}$  and a geometric average between the initial and final values of the sheet diameter, $d\approx \sqrt{d_0 d_{\rm{Max}}}$,  Eq. \ref{eqn:dissip2} reads:

\begin{equation}
\label{eqn:dissip3}
E_B\approx \eta_{\rm{B}}\frac{\pi d_0^2}{6} \frac{\left (d_{\rm{Max}}-d_0\right)^2}{d_{\rm{Max}} t_{\rm{Max}}}
\end{equation}

Here, $\frac{\left (d_{\rm{Max}}-d_0\right)^2}{d_{\rm{Max}} t_{\rm{Max}}}$ can be considered as an apparent velocity, $v_{\rm{app}}$, close to the initial expansion speed of the drop upon impact. Combining Eqs.~\ref{eqn:dtilde} and \ref{eqn:dissip3}  together with the definition of the biaxial extensional Ohnesorge number, one predicts:

 \begin{equation}
\label{eqn:dtilde2}
\tilde{d}=\sqrt{1-\alpha Oh_{\rm{B}}}
\end{equation}

with $\alpha=\frac{2v_{\rm{L}}v_{\rm{app}}}{v_0^2}$ and $v_{\rm{L}}=\sqrt{\frac{\gamma}{\rho do}}$ is the typical velocity of free oscillations of the drop~\cite{Rayleigh:1879vw}. We find that the functional form of Eq.~\ref{eqn:dtilde2} reproduces very nicely the experimental data (fig.~\ref{figure:7}) which implies that $v_{\rm{app}}$ is approximatively constant. The best fit of the data (continuous red line) yields,~for all concentrations, $\alpha=0.06\pm0.02$ where the error bars are used to obtain the two envelops of the data (dashed black lines). This value is compared to the theoretical expectations for the parameter $\alpha$. For Newtonian fluids,  $v_{\rm{app}}=2.3 \pm 1.2\mathrm{m/s}$, thus $\alpha=00.25\pm0.02$, 
 Overall, the fit parameter of the master curve is thus found in reasonable  agreement with the ones calculated using the different experimental quantities, thus justifying the relevance of our approach.

 \section {Conclusion}
Drop impact experiments on repellent surfaces  have been performed with Newtonian fluids and shear-thinning polymer solutions. Two regimes for the maximum expansion diameter of freely expanding sheets have been identified: a capillary regime, where the maximum expansion does not depend on viscosity and a viscous regime where the maximum expansion is reduced with increasing viscosity. We have demonstrated that the dominant source of viscous dissipation is the biaxial extensional deformation during sheet expansion, which consequently controls the maximum expansion of the sheets in the viscous regime. We have provided a scaling prediction of the sheet maximum expansion as a function of the biaxial extensional Ohnesorge number, in good quantitative agreement with our experimental results. For viscoelastic thinning fluids, we have proposed a simple approach to measure the biaxial extensional thinning viscosity based on the maximum  expansion factor of a freely expanding sheet:  the  relevant characteristic thinning viscosity, is simply  given by the viscosity of a Newtonian fluid with the same normalized expansion factor;  it  obeys the behavior of the biaxial extensional viscosity of  polymeric samples as a function of the Weissenberg number in stationary conditions, providing one considers the mean biaxial extensional rate of the sheet during the expansion regime, as the relevant strain rate for the Weissenberg number.  Our approach constitutes a first and crucial step toward the  development of a new class of biaxial extensional rheometry tools based on  drop impact experiments, but needs  further   investigations (in current progress) including  drops of  different  diameters ,  and  different  impact heights .

\begin{acknowledgments}
This work was financially supported by the EU (Marie Sklodowska Curie) ITN Supolen GA N°.607937, the labex NUMEV (ANR-10-LAB-20) and the H2020 Program (Marie Curie Actions) of the European Commission's Innovative Training Networks (ITN) (H2020-MSCA-ITN-2017) under DoDyNet REA Grant Agreement (GA) N°.765811.
We acknowledge, Dr. Sara Lindeblad Wingstrand (DTU) for the discussions concerning extensional rheology, Dr Srishti Arora (Northwestern University), for her help in impact experiments  and Dr. Benoit Loppinet (IESL-FORTH) for fruitful discussions.

\end{acknowledgments}

\bibliography{Ameur1}

\begin{thebibliography}{55}%
\makeatletter
\providecommand \@ifxundefined [1]{%
 \@ifx{#1\undefined}
}%
\providecommand \@ifnum [1]{%
 \ifnum #1\expandafter \@firstoftwo
 \else \expandafter \@secondoftwo
 \fi
}%
\providecommand \@ifx [1]{%
 \ifx #1\expandafter \@firstoftwo
 \else \expandafter \@secondoftwo
 \fi
}%
\providecommand \natexlab [1]{#1}%
\providecommand \enquote  [1]{``#1''}%
\providecommand \bibnamefont  [1]{#1}%
\providecommand \bibfnamefont [1]{#1}%
\providecommand \citenamefont [1]{#1}%
\providecommand \href@noop [0]{\@secondoftwo}%
\providecommand \href [0]{\begingroup \@sanitize@url \@href}%
\providecommand \@href[1]{\@@startlink{#1}\@@href}%
\providecommand \@@href[1]{\endgroup#1\@@endlink}%
\providecommand \@sanitize@url [0]{\catcode `\\12\catcode `\$12\catcode
  `\&12\catcode `\#12\catcode `\^12\catcode `\_12\catcode `\%12\relax}%
\providecommand \@@startlink[1]{}%
\providecommand \@@endlink[0]{}%
\providecommand \url  [0]{\begingroup\@sanitize@url \@url }%
\providecommand \@url [1]{\endgroup\@href {#1}{\urlprefix }}%
\providecommand \urlprefix  [0]{URL }%
\providecommand \Eprint [0]{\href }%
\providecommand \doibase [0]{http://dx.doi.org/}%
\providecommand \selectlanguage [0]{\@gobble}%
\providecommand \bibinfo  [0]{\@secondoftwo}%
\providecommand \bibfield  [0]{\@secondoftwo}%
\providecommand \translation [1]{[#1]}%
\providecommand \BibitemOpen [0]{}%
\providecommand \bibitemStop [0]{}%
\providecommand \bibitemNoStop [0]{.\EOS\space}%
\providecommand \EOS [0]{\spacefactor3000\relax}%
\providecommand \BibitemShut  [1]{\csname bibitem#1\endcsname}%
\let\auto@bib@innerbib\@empty
\bibitem [{\citenamefont {Thoroddsen}\ \emph {et~al.}(2008)\citenamefont
  {Thoroddsen}, \citenamefont {Etoh},\ and\ \citenamefont
  {Takehara}}]{THORODDSEN:2008kl}%
  \BibitemOpen
  \bibfield  {author} {\bibinfo {author} {\bibfnamefont {S~T}\ \bibnamefont
  {Thoroddsen}}, \bibinfo {author} {\bibfnamefont {T~G}\ \bibnamefont {Etoh}},
  \ and\ \bibinfo {author} {\bibfnamefont {K}~\bibnamefont {Takehara}},\
  }\bibfield  {title} {\enquote {\bibinfo {title} {{High-Speed Imaging of Drops
  and Bubbles}},}\ }\href@noop {} {\bibfield  {journal} {\bibinfo  {journal}
  {Annual Review of Fluid Mechanics}\ }\textbf {\bibinfo {volume} {40}},\
  \bibinfo {pages} {257--285} (\bibinfo {year} {2008})}\BibitemShut {NoStop}%
\bibitem [{\citenamefont {Vernay}\ \emph {et~al.}(2015)\citenamefont {Vernay},
  \citenamefont {Ramos},\ and\ \citenamefont {Ligoure}}]{Vernay:2015eg}%
  \BibitemOpen
  \bibfield  {author} {\bibinfo {author} {\bibfnamefont {C}~\bibnamefont
  {Vernay}}, \bibinfo {author} {\bibfnamefont {L}~\bibnamefont {Ramos}}, \ and\
  \bibinfo {author} {\bibfnamefont {C}~\bibnamefont {Ligoure}},\ }\bibfield
  {title} {\enquote {\bibinfo {title} {{Free radially expanding liquid sheet in
  air: time-~and space-resolved measurement of~the~thickness field}},}\
  }\href@noop {} {\bibfield  {journal} {\bibinfo  {journal} {Journal of Fluid
  Mechanics}\ }\textbf {\bibinfo {volume} {764}},\ \bibinfo {pages} {428--444}
  (\bibinfo {year} {2015})}\BibitemShut {NoStop}%
\bibitem [{\citenamefont {Wang}\ and\ \citenamefont
  {Bourouiba}(2017)}]{Wang:2017ju}%
  \BibitemOpen
  \bibfield  {author} {\bibinfo {author} {\bibfnamefont {Y}~\bibnamefont
  {Wang}}\ and\ \bibinfo {author} {\bibfnamefont {L}~\bibnamefont
  {Bourouiba}},\ }\bibfield  {title} {\enquote {\bibinfo {title} {{Drop impact
  on small surfaces: thickness and velocity profiles of the expanding sheet in
  the air}},}\ }\href@noop {} {\bibfield  {journal} {\bibinfo  {journal}
  {Journal of Fluid Mechanics}\ }\textbf {\bibinfo {volume} {814}},\ \bibinfo
  {pages} {510--534} (\bibinfo {year} {2017})}\BibitemShut {NoStop}%
\bibitem [{\citenamefont {Yoon}\ \emph {et~al.}(2007)\citenamefont {Yoon},
  \citenamefont {Jepsen}, \citenamefont {Nissen},\ and\ \citenamefont
  {O{\textquoteright}Hern}}]{Yoon:2007ft}%
  \BibitemOpen
  \bibfield  {author} {\bibinfo {author} {\bibfnamefont {S~S}\ \bibnamefont
  {Yoon}}, \bibinfo {author} {\bibfnamefont {R~A}\ \bibnamefont {Jepsen}},
  \bibinfo {author} {\bibfnamefont {M~R}\ \bibnamefont {Nissen}}, \ and\
  \bibinfo {author} {\bibfnamefont {T~J}\ \bibnamefont
  {O{\textquoteright}Hern}},\ }\bibfield  {title} {\enquote {\bibinfo {title}
  {{Experimental investigation on splashing and nonlinear fingerlike
  instability of large water drops}},}\ }\href@noop {} {\bibfield  {journal}
  {\bibinfo  {journal} {Journal of Fluids and Structures}\ }\textbf {\bibinfo
  {volume} {23}},\ \bibinfo {pages} {101--115} (\bibinfo {year}
  {2007})}\BibitemShut {NoStop}%
\bibitem [{\citenamefont {T}\ and\ \citenamefont
  {Sakakibara}(1998)}]{THORODDSEN:1998kk}%
  \BibitemOpen
  \bibfield  {author} {\bibinfo {author} {\bibfnamefont {Thoroddsen~S}\
  \bibnamefont {T}}\ and\ \bibinfo {author} {\bibfnamefont {J}~\bibnamefont
  {Sakakibara}},\ }\bibfield  {title} {\enquote {\bibinfo {title} {{Evolution
  of the fingering pattern of an impacting drop}},}\ }\href@noop {} {\bibfield
  {journal} {\bibinfo  {journal} {Physics of Fluids}\ }\textbf {\bibinfo
  {volume} {10}},\ \bibinfo {pages} {1359--1374} (\bibinfo {year}
  {1998})}\BibitemShut {NoStop}%
\bibitem [{\citenamefont {Marmanis}\ and\ \citenamefont
  {Thoroddsen}(1996)}]{Marmanis:1996ka}%
  \BibitemOpen
  \bibfield  {author} {\bibinfo {author} {\bibfnamefont {H}~\bibnamefont
  {Marmanis}}\ and\ \bibinfo {author} {\bibfnamefont {S~T}\ \bibnamefont
  {Thoroddsen}},\ }\bibfield  {title} {\enquote {\bibinfo {title} {{Scaling of
  the fingering pattern of an impacting drop}},}\ }\href@noop {} {\bibfield
  {journal} {\bibinfo  {journal} {Physics of Fluids}\ }\textbf {\bibinfo
  {volume} {8}},\ \bibinfo {pages} {1344--1346} (\bibinfo {year}
  {1996})}\BibitemShut {NoStop}%
\bibitem [{\citenamefont {van~der Meer}(2017)}]{vanderMeer:2017bk}%
  \BibitemOpen
  \bibfield  {author} {\bibinfo {author} {\bibfnamefont {Devaraj}\ \bibnamefont
  {van~der Meer}},\ }\bibfield  {title} {\enquote {\bibinfo {title} {{Impact on
  Granular Beds}},}\ }\href@noop {} {\bibfield  {journal} {\bibinfo  {journal}
  {Annual Review of Fluid Mechanics}\ }\textbf {\bibinfo {volume} {49}},\
  \bibinfo {pages} {463--484} (\bibinfo {year} {2017})}\BibitemShut {NoStop}%
\bibitem [{\citenamefont {Josserand}\ and\ \citenamefont
  {Thoroddsen}(2016)}]{Josserand:2016jf}%
  \BibitemOpen
  \bibfield  {author} {\bibinfo {author} {\bibfnamefont {C}~\bibnamefont
  {Josserand}}\ and\ \bibinfo {author} {\bibfnamefont {S~T}\ \bibnamefont
  {Thoroddsen}},\ }\bibfield  {title} {\enquote {\bibinfo {title} {{Drop Impact
  on a Solid Surface}},}\ }\href@noop {} {\bibfield  {journal} {\bibinfo
  {journal} {Annual Review of Fluid Mechanics}\ }\textbf {\bibinfo {volume}
  {48}},\ \bibinfo {pages} {365--391} (\bibinfo {year} {2016})}\BibitemShut
  {NoStop}%
\bibitem [{\citenamefont {Bonn}\ \emph {et~al.}(2009)\citenamefont {Bonn},
  \citenamefont {Eggers}, \citenamefont {Indekeu}, \citenamefont {Meunier},\
  and\ \citenamefont {Rolley}}]{Bonn:2009ha}%
  \BibitemOpen
  \bibfield  {author} {\bibinfo {author} {\bibfnamefont {Daniel}\ \bibnamefont
  {Bonn}}, \bibinfo {author} {\bibfnamefont {Jens}\ \bibnamefont {Eggers}},
  \bibinfo {author} {\bibfnamefont {Joseph}\ \bibnamefont {Indekeu}}, \bibinfo
  {author} {\bibfnamefont {Jacques}\ \bibnamefont {Meunier}}, \ and\ \bibinfo
  {author} {\bibfnamefont {Etienne}\ \bibnamefont {Rolley}},\ }\bibfield
  {title} {\enquote {\bibinfo {title} {{Wetting and spreading}},}\ }\href@noop
  {} {\bibfield  {journal} {\bibinfo  {journal} {Reviews of Modern Physics}\
  }\textbf {\bibinfo {volume} {81}},\ \bibinfo {pages} {739--805} (\bibinfo
  {year} {2009})}\BibitemShut {NoStop}%
\bibitem [{\citenamefont {Boyer}\ \emph {et~al.}(2016)\citenamefont {Boyer},
  \citenamefont {Sandoval-Nava}, \citenamefont {Snoeijer}, \citenamefont
  {Dijksman},\ and\ \citenamefont {Lohse}}]{Boyer:2016gl}%
  \BibitemOpen
  \bibfield  {author} {\bibinfo {author} {\bibfnamefont {Fran{\c c}ois}\
  \bibnamefont {Boyer}}, \bibinfo {author} {\bibfnamefont {Enrique}\
  \bibnamefont {Sandoval-Nava}}, \bibinfo {author} {\bibfnamefont {Jacco~H}\
  \bibnamefont {Snoeijer}}, \bibinfo {author} {\bibfnamefont {J~Frits}\
  \bibnamefont {Dijksman}}, \ and\ \bibinfo {author} {\bibfnamefont {Detlef}\
  \bibnamefont {Lohse}},\ }\bibfield  {title} {\enquote {\bibinfo {title}
  {{Drop impact of shear thickening liquids}},}\ }\href@noop {} {\bibfield
  {journal} {\bibinfo  {journal} {Physical Review Fluids}\ }\textbf {\bibinfo
  {volume} {1}},\ \bibinfo {pages} {013901--9} (\bibinfo {year}
  {2016})}\BibitemShut {NoStop}%
\bibitem [{\citenamefont {Laan}\ \emph {et~al.}(2014)\citenamefont {Laan},
  \citenamefont {de~Bruin}, \citenamefont {Bartolo}, \citenamefont
  {Josserand},\ and\ \citenamefont {Bonn}}]{Laan:2014hj}%
  \BibitemOpen
  \bibfield  {author} {\bibinfo {author} {\bibfnamefont {Nick}\ \bibnamefont
  {Laan}}, \bibinfo {author} {\bibfnamefont {Karla~G}\ \bibnamefont
  {de~Bruin}}, \bibinfo {author} {\bibfnamefont {Denis}\ \bibnamefont
  {Bartolo}}, \bibinfo {author} {\bibfnamefont {Christophe}\ \bibnamefont
  {Josserand}}, \ and\ \bibinfo {author} {\bibfnamefont {Daniel}\ \bibnamefont
  {Bonn}},\ }\bibfield  {title} {\enquote {\bibinfo {title} {{Maximum Diameter
  of Impacting Liquid Droplets}},}\ }\href@noop {} {\bibfield  {journal}
  {\bibinfo  {journal} {Physical Review Applied}\ }\textbf {\bibinfo {volume}
  {2}},\ \bibinfo {pages} {863--7} (\bibinfo {year} {2014})}\BibitemShut
  {NoStop}%
\bibitem [{\citenamefont {An}\ and\ \citenamefont {Lee}(2012)}]{An:2012gc}%
  \BibitemOpen
  \bibfield  {author} {\bibinfo {author} {\bibfnamefont {Sang~Mo}\ \bibnamefont
  {An}}\ and\ \bibinfo {author} {\bibfnamefont {Sang~Yong}\ \bibnamefont
  {Lee}},\ }\bibfield  {title} {\enquote {\bibinfo {title} {{Maximum spreading
  of a shear-thinning liquid drop impacting on dry solid surfaces}},}\
  }\href@noop {} {\bibfield  {journal} {\bibinfo  {journal} {Experimental
  Thermal and Fluid Science}\ }\textbf {\bibinfo {volume} {38}},\ \bibinfo
  {pages} {140--148} (\bibinfo {year} {2012})}\BibitemShut {NoStop}%
\bibitem [{\citenamefont {Cooper-White}\ \emph {et~al.}(2002)\citenamefont
  {Cooper-White}, \citenamefont {Crooks},\ and\ \citenamefont
  {Boger}}]{CooperWhite:2002vt}%
  \BibitemOpen
  \bibfield  {author} {\bibinfo {author} {\bibfnamefont {J~J}\ \bibnamefont
  {Cooper-White}}, \bibinfo {author} {\bibfnamefont {R~C}\ \bibnamefont
  {Crooks}}, \ and\ \bibinfo {author} {\bibfnamefont {D~V}\ \bibnamefont
  {Boger}},\ }\bibfield  {title} {\enquote {\bibinfo {title} {{A drop impact
  study of worm-like viscoelastic surfactant solutions}},}\ }\href@noop {}
  {\bibfield  {journal} {\bibinfo  {journal} {Colloids and Surfaces
  a-Physicochemical and Engineering Aspects}\ }\textbf {\bibinfo {volume}
  {210}},\ \bibinfo {pages} {105--123} (\bibinfo {year} {2002})}\BibitemShut
  {NoStop}%
\bibitem [{\citenamefont {Crooks}\ and\ \citenamefont
  {Boger}(2000)}]{Crooks:2000bt}%
  \BibitemOpen
  \bibfield  {author} {\bibinfo {author} {\bibfnamefont {Regan}\ \bibnamefont
  {Crooks}}\ and\ \bibinfo {author} {\bibfnamefont {David~V}\ \bibnamefont
  {Boger}},\ }\bibfield  {title} {\enquote {\bibinfo {title} {{Influence of
  fluid elasticity on drops impacting on dry surfaces}},}\ }\href@noop {}
  {\bibfield  {journal} {\bibinfo  {journal} {Journal of Rheology}\ }\textbf
  {\bibinfo {volume} {44}},\ \bibinfo {pages} {973--996} (\bibinfo {year}
  {2000})}\BibitemShut {NoStop}%
\bibitem [{\citenamefont {Izbassarov}\ and\ \citenamefont
  {Muradoglu}(2016)}]{Izbassarov:jq}%
  \BibitemOpen
  \bibfield  {author} {\bibinfo {author} {\bibfnamefont {D}~\bibnamefont
  {Izbassarov}}\ and\ \bibinfo {author} {\bibfnamefont {M}~\bibnamefont
  {Muradoglu}},\ }\bibfield  {title} {\enquote {\bibinfo {title} {{Effects of
  viscoelasticity on drop impact and spreading on a solid surface}},}\
  }\href@noop {} {\bibfield  {journal} {\bibinfo  {journal} {Physical Review
  Fluids}\ }\textbf {\bibinfo {volume} {1}},\ \bibinfo {pages} {023302}
  (\bibinfo {year} {2016})}\BibitemShut {NoStop}%
\bibitem [{\citenamefont {Rozhkov}\ \emph {et~al.}(2006)\citenamefont
  {Rozhkov}, \citenamefont {Prunet-Foch},\ and\ \citenamefont
  {Vignes-Adler}}]{Rozhkov:2006ic}%
  \BibitemOpen
  \bibfield  {author} {\bibinfo {author} {\bibfnamefont {Aleksey}\ \bibnamefont
  {Rozhkov}}, \bibinfo {author} {\bibfnamefont {Bernard}\ \bibnamefont
  {Prunet-Foch}}, \ and\ \bibinfo {author} {\bibfnamefont {Mich{\`e}le}\
  \bibnamefont {Vignes-Adler}},\ }\bibfield  {title} {\enquote {\bibinfo
  {title} {{Dynamics and disintegration of drops of polymeric liquids}},}\
  }\href@noop {} {\bibfield  {journal} {\bibinfo  {journal} {Journal of
  Non-Newtonian Fluid Mechanics}\ }\textbf {\bibinfo {volume} {134}},\ \bibinfo
  {pages} {44--55} (\bibinfo {year} {2006})}\BibitemShut {NoStop}%
\bibitem [{\citenamefont {Huh}\ \emph {et~al.}(2015)\citenamefont {Huh},
  \citenamefont {Jung}, \citenamefont {Seo},\ and\ \citenamefont
  {Lee}}]{Huh:2015jq}%
  \BibitemOpen
  \bibfield  {author} {\bibinfo {author} {\bibfnamefont {Hyung~Kyu}\
  \bibnamefont {Huh}}, \bibinfo {author} {\bibfnamefont {Sungjune}\
  \bibnamefont {Jung}}, \bibinfo {author} {\bibfnamefont {Kyung~Won}\
  \bibnamefont {Seo}}, \ and\ \bibinfo {author} {\bibfnamefont {Sang~Joon}\
  \bibnamefont {Lee}},\ }\bibfield  {title} {\enquote {\bibinfo {title} {{Role
  of polymer concentration and molecular weight on the rebounding behaviors of
  polymer solution droplet impacting on hydrophobic surfaces}},}\ }\href@noop
  {} {\bibfield  {journal} {\bibinfo  {journal} {Microfluidics and
  Nanofluidics}\ ,\ \bibinfo {pages} {1--13}} (\bibinfo {year}
  {2015})}\BibitemShut {NoStop}%
\bibitem [{\citenamefont {Luu}\ and\ \citenamefont
  {Forterre}(2009)}]{Luu:2009jr}%
  \BibitemOpen
  \bibfield  {author} {\bibinfo {author} {\bibfnamefont {Li-Hua}\ \bibnamefont
  {Luu}}\ and\ \bibinfo {author} {\bibfnamefont {Yo{\"e}l}\ \bibnamefont
  {Forterre}},\ }\bibfield  {title} {\enquote {\bibinfo {title} {{Drop impact
  of yield-stress fluids}},}\ }\href@noop {} {\bibfield  {journal} {\bibinfo
  {journal} {Journal of Fluid Mechanics}\ }\textbf {\bibinfo {volume} {632}},\
  \bibinfo {pages} {301--327} (\bibinfo {year} {2009})}\BibitemShut {NoStop}%
\bibitem [{\citenamefont {Arora}\ \emph {et~al.}(2016)\citenamefont {Arora},
  \citenamefont {Ligoure},\ and\ \citenamefont {Ramos}}]{Arora:2016bu}%
  \BibitemOpen
  \bibfield  {author} {\bibinfo {author} {\bibfnamefont {Srishti}\ \bibnamefont
  {Arora}}, \bibinfo {author} {\bibfnamefont {Christian}\ \bibnamefont
  {Ligoure}}, \ and\ \bibinfo {author} {\bibfnamefont {Laurence}\ \bibnamefont
  {Ramos}},\ }\bibfield  {title} {\enquote {\bibinfo {title} {{Interplay
  between viscosity and elasticity in freely expanding liquid sheets}},}\
  }\href@noop {} {\bibfield  {journal} {\bibinfo  {journal} {Phys. Rev.
  Fluids}\ }\textbf {\bibinfo {volume} {1}},\ \bibinfo {pages} {083302--15}
  (\bibinfo {year} {2016})}\BibitemShut {NoStop}%
\bibitem [{\citenamefont {Arora}\ \emph {et~al.}(2018)\citenamefont {Arora},
  \citenamefont {Fromental}, \citenamefont {Mora}, \citenamefont {Phou},
  \citenamefont {Ramos},\ and\ \citenamefont {Ligoure}}]{Arora:2018ei}%
  \BibitemOpen
  \bibfield  {author} {\bibinfo {author} {\bibfnamefont {S}~\bibnamefont
  {Arora}}, \bibinfo {author} {\bibfnamefont {J~M}\ \bibnamefont {Fromental}},
  \bibinfo {author} {\bibfnamefont {S}~\bibnamefont {Mora}}, \bibinfo {author}
  {\bibfnamefont {Ty}~\bibnamefont {Phou}}, \bibinfo {author} {\bibfnamefont
  {L}~\bibnamefont {Ramos}}, \ and\ \bibinfo {author} {\bibfnamefont
  {C}~\bibnamefont {Ligoure}},\ }\bibfield  {title} {\enquote {\bibinfo {title}
  {{Impact of Beads and Drops on a Repellent Solid Surface: A Unified
  Description}},}\ }\href@noop {} {\bibfield  {journal} {\bibinfo  {journal}
  {Physical Review Letters}\ }\textbf {\bibinfo {volume} {120}},\ \bibinfo
  {pages} {148003} (\bibinfo {year} {2018})}\BibitemShut {NoStop}%
\bibitem [{\citenamefont {Rioboo}\ \emph {et~al.}(2002)\citenamefont {Rioboo},
  \citenamefont {Marengo},\ and\ \citenamefont {Tropea}}]{Rioboo:2002gk}%
  \BibitemOpen
  \bibfield  {author} {\bibinfo {author} {\bibfnamefont {R}~\bibnamefont
  {Rioboo}}, \bibinfo {author} {\bibfnamefont {M}~\bibnamefont {Marengo}}, \
  and\ \bibinfo {author} {\bibfnamefont {C}~\bibnamefont {Tropea}},\ }\bibfield
   {title} {\enquote {\bibinfo {title} {{Time evolution of liquid drop impact
  onto solid, dry surfaces}},}\ }\href@noop {} {\bibfield  {journal} {\bibinfo
  {journal} {Experiments in Fluids}\ }\textbf {\bibinfo {volume} {33}},\
  \bibinfo {pages} {112--124} (\bibinfo {year} {2002})}\BibitemShut {NoStop}%
\bibitem [{\citenamefont {Moreira}\ \emph {et~al.}(2007)\citenamefont
  {Moreira}, \citenamefont {Moita}, \citenamefont {Cossali}, \citenamefont
  {Marengo},\ and\ \citenamefont {Santini}}]{Moreira:2007gh}%
  \BibitemOpen
  \bibfield  {author} {\bibinfo {author} {\bibfnamefont {A~L~N}\ \bibnamefont
  {Moreira}}, \bibinfo {author} {\bibfnamefont {A~S}\ \bibnamefont {Moita}},
  \bibinfo {author} {\bibfnamefont {E}~\bibnamefont {Cossali}}, \bibinfo
  {author} {\bibfnamefont {M}~\bibnamefont {Marengo}}, \ and\ \bibinfo {author}
  {\bibfnamefont {M}~\bibnamefont {Santini}},\ }\bibfield  {title} {\enquote
  {\bibinfo {title} {{Secondary atomization of water and isooctane drops
  impinging on tilted heated surfaces}},}\ }\href@noop {} {\bibfield  {journal}
  {\bibinfo  {journal} {Experiments in Fluids}\ }\textbf {\bibinfo {volume}
  {43}},\ \bibinfo {pages} {297--313} (\bibinfo {year} {2007})}\BibitemShut
  {NoStop}%
\bibitem [{\citenamefont {Lee}\ \emph {et~al.}(2010)\citenamefont {Lee},
  \citenamefont {Chang},\ and\ \citenamefont {Kim}}]{Lee:2010bt}%
  \BibitemOpen
  \bibfield  {author} {\bibinfo {author} {\bibfnamefont {Minhee}\ \bibnamefont
  {Lee}}, \bibinfo {author} {\bibfnamefont {Young~Soo}\ \bibnamefont {Chang}},
  \ and\ \bibinfo {author} {\bibfnamefont {Ho-Young}\ \bibnamefont {Kim}},\
  }\bibfield  {title} {\enquote {\bibinfo {title} {{Drop impact on microwetting
  patterned surfaces}},}\ }\href@noop {} {\bibfield  {journal} {\bibinfo
  {journal} {Physics of Fluids}\ }\textbf {\bibinfo {volume} {22}},\ \bibinfo
  {pages} {072101--8} (\bibinfo {year} {2010})}\BibitemShut {NoStop}%
\bibitem [{\citenamefont {German}\ and\ \citenamefont
  {Bertola}(2009)}]{German:2009hp}%
  \BibitemOpen
  \bibfield  {author} {\bibinfo {author} {\bibfnamefont {G}~\bibnamefont
  {German}}\ and\ \bibinfo {author} {\bibfnamefont {V}~\bibnamefont
  {Bertola}},\ }\bibfield  {title} {\enquote {\bibinfo {title} {{Impact of
  shear-thinning and yield-stress drops on solid substrates}},}\ }\href@noop {}
  {\bibfield  {journal} {\bibinfo  {journal} {Journal of Physics-Condensed
  Matter}\ }\textbf {\bibinfo {volume} {21}},\ \bibinfo {pages} {375111--17}
  (\bibinfo {year} {2009})}\BibitemShut {NoStop}%
\bibitem [{\citenamefont {Lee}\ \emph {et~al.}(2016)\citenamefont {Lee},
  \citenamefont {Derome}, \citenamefont {Guyer},\ and\ \citenamefont
  {Carmeliet}}]{Lee:2016ee}%
  \BibitemOpen
  \bibfield  {author} {\bibinfo {author} {\bibfnamefont {Jae~Bong}\
  \bibnamefont {Lee}}, \bibinfo {author} {\bibfnamefont {Dominique}\
  \bibnamefont {Derome}}, \bibinfo {author} {\bibfnamefont {Robert}\
  \bibnamefont {Guyer}}, \ and\ \bibinfo {author} {\bibfnamefont {Jan}\
  \bibnamefont {Carmeliet}},\ }\bibfield  {title} {\enquote {\bibinfo {title}
  {{Modeling the Maximum Spreading of Liquid Droplets Impacting Wetting and
  Nonwetting Surfaces}},}\ }\href@noop {} {\bibfield  {journal} {\bibinfo
  {journal} {Langmuir}\ }\textbf {\bibinfo {volume} {32}},\ \bibinfo {pages}
  {1299--1308} (\bibinfo {year} {2016})}\BibitemShut {NoStop}%
\bibitem [{\citenamefont {Ukiwe}\ and\ \citenamefont
  {Kwok}(2005)}]{Ukiwe:2005jw}%
  \BibitemOpen
  \bibfield  {author} {\bibinfo {author} {\bibfnamefont {Chijioke}\
  \bibnamefont {Ukiwe}}\ and\ \bibinfo {author} {\bibfnamefont {Daniel~Y}\
  \bibnamefont {Kwok}},\ }\bibfield  {title} {\enquote {\bibinfo {title} {{On
  the Maximum Spreading Diameter of Impacting Droplets on Well-Prepared Solid
  Surfaces}},}\ }\href@noop {} {\bibfield  {journal} {\bibinfo  {journal}
  {Langmuir}\ }\textbf {\bibinfo {volume} {21}},\ \bibinfo {pages} {666--673}
  (\bibinfo {year} {2005})}\BibitemShut {NoStop}%
\bibitem [{\citenamefont {Roux}\ and\ \citenamefont
  {Cooper-White}(2004)}]{Roux:2004dx}%
  \BibitemOpen
  \bibfield  {author} {\bibinfo {author} {\bibfnamefont {D~C~D}\ \bibnamefont
  {Roux}}\ and\ \bibinfo {author} {\bibfnamefont {J~J}\ \bibnamefont
  {Cooper-White}},\ }\bibfield  {title} {\enquote {\bibinfo {title} {{Dynamics
  of water spreading on a glass surface}},}\ }\href@noop {} {\bibfield
  {journal} {\bibinfo  {journal} {Journal of Colloid and Interface Science}\
  }\textbf {\bibinfo {volume} {277}},\ \bibinfo {pages} {424--436} (\bibinfo
  {year} {2004})}\BibitemShut {NoStop}%
\bibitem [{\citenamefont {Rozhkov}\ \emph {et~al.}(2004)\citenamefont
  {Rozhkov}, \citenamefont {Prunet-Foch},\ and\ \citenamefont
  {Vignes-Adler}}]{Rozhkov:2004in}%
  \BibitemOpen
  \bibfield  {author} {\bibinfo {author} {\bibfnamefont {A}~\bibnamefont
  {Rozhkov}}, \bibinfo {author} {\bibfnamefont {B}~\bibnamefont {Prunet-Foch}},
  \ and\ \bibinfo {author} {\bibfnamefont {M}~\bibnamefont {Vignes-Adler}},\
  }\bibfield  {title} {\enquote {\bibinfo {title} {{Dynamics of a liquid
  lamella resulting from the impact of a water drop on a small target}},}\
  }\href@noop {} {\bibfield  {journal} {\bibinfo  {journal} {Proc. R. Soc.
  Lond.}\ }\textbf {\bibinfo {volume} {460}},\ \bibinfo {pages} {2681--2704}
  (\bibinfo {year} {2004})}\BibitemShut {NoStop}%
\bibitem [{\citenamefont {Villermaux}\ and\ \citenamefont
  {Bossa}(2011)}]{Villermaux:2011ff}%
  \BibitemOpen
  \bibfield  {author} {\bibinfo {author} {\bibfnamefont {E}~\bibnamefont
  {Villermaux}}\ and\ \bibinfo {author} {\bibfnamefont {B}~\bibnamefont
  {Bossa}},\ }\bibfield  {title} {\enquote {\bibinfo {title} {{Drop
  fragmentation on impact}},}\ }\href@noop {} {\bibfield  {journal} {\bibinfo
  {journal} {Journal of Fluid Mechanics}\ }\textbf {\bibinfo {volume} {668}},\
  \bibinfo {pages} {412--435} (\bibinfo {year} {2011})}\BibitemShut {NoStop}%
\bibitem [{\citenamefont {Lejeune}\ \emph {et~al.}(2018)\citenamefont
  {Lejeune}, \citenamefont {Gilet},\ and\ \citenamefont
  {Bourouiba}}]{Lejeune:2018da}%
  \BibitemOpen
  \bibfield  {author} {\bibinfo {author} {\bibfnamefont {S}~\bibnamefont
  {Lejeune}}, \bibinfo {author} {\bibfnamefont {T}~\bibnamefont {Gilet}}, \
  and\ \bibinfo {author} {\bibfnamefont {L}~\bibnamefont {Bourouiba}},\
  }\bibfield  {title} {\enquote {\bibinfo {title} {{Edge effect: Liquid sheet
  and droplets formed by drop impact close to an edge}},}\ }\href@noop {}
  {\bibfield  {journal} {\bibinfo  {journal} {Physical Review Fluids}\ }\textbf
  {\bibinfo {volume} {3}},\ \bibinfo {pages} {1--32} (\bibinfo {year}
  {2018})}\BibitemShut {NoStop}%
\bibitem [{\citenamefont {Richard}\ \emph {et~al.}(2002)\citenamefont
  {Richard}, \citenamefont {Clanet},\ and\ \citenamefont
  {Qu{\'e}r{\'e}}}]{Richard2002}%
  \BibitemOpen
  \bibfield  {author} {\bibinfo {author} {\bibfnamefont {Denis}\ \bibnamefont
  {Richard}}, \bibinfo {author} {\bibfnamefont {Christophe}\ \bibnamefont
  {Clanet}}, \ and\ \bibinfo {author} {\bibfnamefont {David}\ \bibnamefont
  {Qu{\'e}r{\'e}}},\ }\bibfield  {title} {\enquote {\bibinfo {title} {Surface
  phenomena: Contact time of a bouncing drop},}\ }\href@noop {} {\bibfield
  {journal} {\bibinfo  {journal} {Nature}\ }\textbf {\bibinfo {volume} {417}},\
  \bibinfo {pages} {811--811} (\bibinfo {year} {2002})}\BibitemShut {NoStop}%
\bibitem [{\citenamefont {Wachters}\ \emph {et~al.}(1966)\citenamefont
  {Wachters}, \citenamefont {Smulders}, \citenamefont {Vermeulen},\ and\
  \citenamefont {HC}}]{Wachters:1966vy}%
  \BibitemOpen
  \bibfield  {author} {\bibinfo {author} {\bibfnamefont {LHJ}\ \bibnamefont
  {Wachters}}, \bibinfo {author} {\bibfnamefont {L}~\bibnamefont {Smulders}},
  \bibinfo {author} {\bibfnamefont {J~R}\ \bibnamefont {Vermeulen}}, \ and\
  \bibinfo {author} {\bibfnamefont {Kleigweig}\ \bibnamefont {HC}},\ }\bibfield
   {title} {\enquote {\bibinfo {title} {{The heat transfer from a hot wall to
  impinging mist droplets in the spheroidal state}},}\ }\href@noop {}
  {\bibfield  {journal} {\bibinfo  {journal} {Chemical Engineering Science}\
  }\textbf {\bibinfo {volume} {21}},\ \bibinfo {pages} {1231--1238} (\bibinfo
  {year} {1966})}\BibitemShut {NoStop}%
\bibitem [{\citenamefont {Biance}\ \emph {et~al.}(2003)\citenamefont {Biance},
  \citenamefont {Clanet},\ and\ \citenamefont {Qu{\'e}r{\'e}}}]{Biance2003}%
  \BibitemOpen
  \bibfield  {author} {\bibinfo {author} {\bibfnamefont {Anne-Laure}\
  \bibnamefont {Biance}}, \bibinfo {author} {\bibfnamefont {Christophe}\
  \bibnamefont {Clanet}}, \ and\ \bibinfo {author} {\bibfnamefont {David}\
  \bibnamefont {Qu{\'e}r{\'e}}},\ }\href@noop {} {\bibfield  {journal}
  {\bibinfo  {journal} {Phys. Fluids}\ }\textbf {\bibinfo {volume} {15}},\
  \bibinfo {pages} {1632} (\bibinfo {year} {2003})}\BibitemShut {NoStop}%
\bibitem [{\citenamefont {Antonini}\ \emph {et~al.}(2013)\citenamefont
  {Antonini}, \citenamefont {Bernagozzi}, \citenamefont {Jung}, \citenamefont
  {Poulikakos},\ and\ \citenamefont {Marengo}}]{Antonini2013}%
  \BibitemOpen
  \bibfield  {author} {\bibinfo {author} {\bibfnamefont {C}~\bibnamefont
  {Antonini}}, \bibinfo {author} {\bibfnamefont {I}~\bibnamefont {Bernagozzi}},
  \bibinfo {author} {\bibfnamefont {S}~\bibnamefont {Jung}}, \bibinfo {author}
  {\bibfnamefont {D}~\bibnamefont {Poulikakos}}, \ and\ \bibinfo {author}
  {\bibfnamefont {M}~\bibnamefont {Marengo}},\ }\bibfield  {title} {\enquote
  {\bibinfo {title} {Water drops dancing on ice: how sublimation leads to drop
  rebound},}\ }\href@noop {} {\bibfield  {journal} {\bibinfo  {journal} {Phys.
  Rev. Lett.}\ }\textbf {\bibinfo {volume} {111}},\ \bibinfo {pages} {014501}
  (\bibinfo {year} {2013})}\BibitemShut {NoStop}%
\bibitem [{\citenamefont {Pack}\ \emph {et~al.}(2019)\citenamefont {Pack},
  \citenamefont {Yang}, \citenamefont {Perazzo}, \citenamefont {Qin},\ and\
  \citenamefont {Stone}}]{Pack:2019hn}%
  \BibitemOpen
  \bibfield  {author} {\bibinfo {author} {\bibfnamefont {Min~Y}\ \bibnamefont
  {Pack}}, \bibinfo {author} {\bibfnamefont {Angela}\ \bibnamefont {Yang}},
  \bibinfo {author} {\bibfnamefont {Antonio}\ \bibnamefont {Perazzo}}, \bibinfo
  {author} {\bibfnamefont {Boyang}\ \bibnamefont {Qin}}, \ and\ \bibinfo
  {author} {\bibfnamefont {Howard~A}\ \bibnamefont {Stone}},\ }\bibfield
  {title} {\enquote {\bibinfo {title} {{Role of extensional rheology on droplet
  bouncing}},}\ }\href@noop {} {\bibfield  {journal} {\bibinfo  {journal}
  {Physical Review Fluids}\ }\textbf {\bibinfo {volume} {4}},\ \bibinfo {pages}
  {123603} (\bibinfo {year} {2019})}\BibitemShut {NoStop}%
\bibitem [{\citenamefont {Cao}\ and\ \citenamefont {Kim}(1994)}]{Cao:1994kv}%
  \BibitemOpen
  \bibfield  {author} {\bibinfo {author} {\bibfnamefont {B~H}\ \bibnamefont
  {Cao}}\ and\ \bibinfo {author} {\bibfnamefont {Mahn~Won}\ \bibnamefont
  {Kim}},\ }\bibfield  {title} {\enquote {\bibinfo {title} {{Molecular weight
  dependence of the surface tension of aqueous poly(ethylene oxide)
  solutions}},}\ }\href@noop {} {\bibfield  {journal} {\bibinfo  {journal}
  {Faraday Discussions}\ }\textbf {\bibinfo {volume} {98}},\ \bibinfo {pages}
  {245--8} (\bibinfo {year} {1994})}\BibitemShut {NoStop}%
\bibitem [{\citenamefont {Crisp}\ \emph {et~al.}(1987)\citenamefont {Crisp},
  \citenamefont {E},\ and\ \citenamefont {Tiedeman}}]{ISI:A1987G694200038}%
  \BibitemOpen
  \bibfield  {author} {\bibinfo {author} {\bibfnamefont {A}~\bibnamefont
  {Crisp}}, \bibinfo {author} {\bibfnamefont {Dejuan}\ \bibnamefont {E}}, \
  and\ \bibinfo {author} {\bibfnamefont {J}~\bibnamefont {Tiedeman}},\
  }\bibfield  {title} {\enquote {\bibinfo {title} {{Effect of silicone oil
  viscosity on emulsification }},}\ }\href@noop {} {\bibfield  {journal}
  {\bibinfo  {journal} {Archives of ophtalmology}\ }\textbf {\bibinfo {volume}
  {105}},\ \bibinfo {pages} {546--550} (\bibinfo {year} {1987})}\BibitemShut
  {NoStop}%
\bibitem [{\citenamefont {Chen}\ and\ \citenamefont
  {Bertola}(2016)}]{Chen:2016fo}%
  \BibitemOpen
  \bibfield  {author} {\bibinfo {author} {\bibfnamefont {Simeng}\ \bibnamefont
  {Chen}}\ and\ \bibinfo {author} {\bibfnamefont {Volfango}\ \bibnamefont
  {Bertola}},\ }\bibfield  {title} {\enquote {\bibinfo {title} {{The impact of
  viscoplastic drops on a heated surface in the Leidenfrost regime}},}\
  }\href@noop {} {\bibfield  {journal} {\bibinfo  {journal} {Soft Matter}\
  }\textbf {\bibinfo {volume} {12}},\ \bibinfo {pages} {7624--7631} (\bibinfo
  {year} {2016})}\BibitemShut {NoStop}%
\bibitem [{\citenamefont {M}\ \emph {et~al.}(1994)\citenamefont {M},
  \citenamefont {Dekee},\ and\ \citenamefont {Carreau}}]{Ortiz:1994kh}%
  \BibitemOpen
  \bibfield  {author} {\bibinfo {author} {\bibfnamefont {Ortiz}\ \bibnamefont
  {M}}, \bibinfo {author} {\bibfnamefont {D}~\bibnamefont {Dekee}}, \ and\
  \bibinfo {author} {\bibfnamefont {PJ}~\bibnamefont {Carreau}},\ }\bibfield
  {title} {\enquote {\bibinfo {title} {{Rheology of concentrated polyethylene
  oxide solutions}},}\ }\href {\doibase {10.1122/1.550472}} {\bibfield
  {journal} {\bibinfo  {journal} {Journal of Rheology}\ }\textbf {\bibinfo
  {volume} {38}},\ \bibinfo {pages} {519--539} (\bibinfo {year}
  {1994})}\BibitemShut {NoStop}%
\bibitem [{\citenamefont {Rubinstein}\ and\ \citenamefont
  {Colby}(2003)}]{Rubinstein:2003bk}%
  \BibitemOpen
  \bibfield  {author} {\bibinfo {author} {\bibfnamefont {M}~\bibnamefont
  {Rubinstein}}\ and\ \bibinfo {author} {\bibfnamefont {R.~H.}\ \bibnamefont
  {Colby}},\ }\href@noop {} {\emph {\bibinfo {title} {Polymer Physics}}}\
  (\bibinfo  {publisher} {Oxford University Press},\ \bibinfo {year}
  {2003})\BibitemShut {NoStop}%
\bibitem [{\citenamefont {Cox}\ and\ \citenamefont {Merz}(1958)}]{Cox:1958jn}%
  \BibitemOpen
  \bibfield  {author} {\bibinfo {author} {\bibfnamefont {W~P}\ \bibnamefont
  {Cox}}\ and\ \bibinfo {author} {\bibfnamefont {E~H}\ \bibnamefont {Merz}},\
  }\bibfield  {title} {\enquote {\bibinfo {title} {{Correlation of dynamic and
  steady flow viscosities}},}\ }\href@noop {} {\bibfield  {journal} {\bibinfo
  {journal} {Journal of Polymer Science}\ }\textbf {\bibinfo {volume} {28}},\
  \bibinfo {pages} {619--622} (\bibinfo {year} {1958})}\BibitemShut {NoStop}%
\bibitem [{\citenamefont {Cross}(1965)}]{Cross:1965bm}%
  \BibitemOpen
  \bibfield  {author} {\bibinfo {author} {\bibfnamefont {MM}~\bibnamefont
  {Cross}},\ }\bibfield  {title} {\enquote {\bibinfo {title} {{Rheology of
  non-Newtonian fluids: a new flow equation for pseudoplastic systems}},}\
  }\href@noop {} {\bibfield  {journal} {\bibinfo  {journal} {Journal of Colloid
  Science}\ }\textbf {\bibinfo {volume} {30}},\ \bibinfo {pages} {417--437}
  (\bibinfo {year} {1965})}\BibitemShut {NoStop}%
\bibitem [{\citenamefont {Macosko}(1994)}]{Macosko:1994bk}%
  \BibitemOpen
  \bibfield  {author} {\bibinfo {author} {\bibfnamefont {C.~W.}\ \bibnamefont
  {Macosko}},\ }\href@noop {} {\emph {\bibinfo {title} {Rheology: Principles,
  Measurements and Applications}}}\ (\bibinfo  {publisher} {Wiley-VCH},\
  \bibinfo {year} {1994})\BibitemShut {NoStop}%
\bibitem [{\citenamefont {Oberstar}(1951)}]{Oberstar:1951im}%
  \BibitemOpen
  \bibfield  {author} {\bibinfo {author} {\bibfnamefont {J~B Segur Helen~E}\
  \bibnamefont {Oberstar}},\ }\bibfield  {title} {\enquote {\bibinfo {title}
  {{Viscosity of Glycerol and Its Aqueous Solutions}},}\ }\href@noop {}
  {\bibfield  {journal} {\bibinfo  {journal} {Ind. Eng. Chem.}\ }\textbf
  {\bibinfo {volume} {43}},\ \bibinfo {pages} {2117--2120} (\bibinfo {year}
  {1951})}\BibitemShut {NoStop}%
\bibitem [{\citenamefont {Maerker}\ and\ \citenamefont
  {Schowalter}(1974)}]{Maerker:1974jx}%
  \BibitemOpen
  \bibfield  {author} {\bibinfo {author} {\bibfnamefont {J~M}\ \bibnamefont
  {Maerker}}\ and\ \bibinfo {author} {\bibfnamefont {W~R}\ \bibnamefont
  {Schowalter}},\ }\bibfield  {title} {\enquote {\bibinfo {title} {{Biaxial
  extension of an elastic liquid}},}\ }\href@noop {} {\bibfield  {journal}
  {\bibinfo  {journal} {Rheologica acta}\ }\textbf {\bibinfo {volume} {13}},\
  \bibinfo {pages} {627--638} (\bibinfo {year} {1974})}\BibitemShut {NoStop}%
\bibitem [{\citenamefont {Joye}\ \emph {et~al.}(1972)\citenamefont {Joye},
  \citenamefont {Poehlein},\ and\ \citenamefont {Denson}}]{Joye:1972pd}%
  \BibitemOpen
  \bibfield  {author} {\bibinfo {author} {\bibfnamefont {D.D}\ \bibnamefont
  {Joye}}, \bibinfo {author} {\bibfnamefont {G.W.}\ \bibnamefont {Poehlein}}, \
  and\ \bibinfo {author} {\bibfnamefont {C.D.}\ \bibnamefont {Denson}},\
  }\bibfield  {title} {\enquote {\bibinfo {title} {A bubble inflation technique
  for the measurement of viscoelastic properties in equal biaxial extensional
  flow},}\ }\href@noop {} {\bibfield  {journal} {\bibinfo  {journal}
  {Transactions of the Society of Rheology}\ }\textbf {\bibinfo {volume}
  {53}},\ \bibinfo {pages} {122--131} (\bibinfo {year} {1972})}\BibitemShut
  {NoStop}%
\bibitem [{\citenamefont {Venerus}\ \emph {et~al.}(2010)\citenamefont
  {Venerus}, \citenamefont {Shiu}, \citenamefont {Kashyap},\ and\ \citenamefont
  {Hosttetler}}]{Venerus:2010jl}%
  \BibitemOpen
  \bibfield  {author} {\bibinfo {author} {\bibfnamefont {D.~C.}\ \bibnamefont
  {Venerus}}, \bibinfo {author} {\bibfnamefont {T.~Y.}\ \bibnamefont {Shiu}},
  \bibinfo {author} {\bibfnamefont {T.}~\bibnamefont {Kashyap}}, \ and\
  \bibinfo {author} {\bibfnamefont {J.}~\bibnamefont {Hosttetler}},\ }\bibfield
   {title} {\enquote {\bibinfo {title} {{Continuous lubricated squeezing flow:
  A novel technique for equibiaxial elongational viscosity measurements on
  polymer melts}},}\ }\href@noop {} {\bibfield  {journal} {\bibinfo  {journal}
  {Journal of Rheology}\ }\textbf {\bibinfo {volume} {54}},\ \bibinfo {pages}
  {1083--1095} (\bibinfo {year} {2010})}\BibitemShut {NoStop}%
\bibitem [{\citenamefont {Johnson}\ \emph {et~al.}(2016)\citenamefont
  {Johnson}, \citenamefont {Murphy}, \citenamefont {Ekins}, \citenamefont
  {Hanley},\ and\ \citenamefont {Jerrams}}]{Johnson2016:bn}%
  \BibitemOpen
  \bibfield  {author} {\bibinfo {author} {\bibfnamefont {Mark}\ \bibnamefont
  {Johnson}}, \bibinfo {author} {\bibfnamefont {Niall}\ \bibnamefont {Murphy}},
  \bibinfo {author} {\bibfnamefont {Ray}\ \bibnamefont {Ekins}}, \bibinfo
  {author} {\bibfnamefont {John}\ \bibnamefont {Hanley}}, \ and\ \bibinfo
  {author} {\bibfnamefont {Stephen}\ \bibnamefont {Jerrams}},\ }\bibfield
  {title} {\enquote {\bibinfo {title} {{Equi-biaxial fatigue testing of EPM
  utilising bubble inflation}},}\ }\href@noop {} {\bibfield  {journal}
  {\bibinfo  {journal} {Polymer testing}\ }\textbf {\bibinfo {volume} {53}},\
  \bibinfo {pages} {122--131} (\bibinfo {year} {2016})}\BibitemShut {NoStop}%
\bibitem [{\citenamefont {Huang}\ and\ \citenamefont
  {Kokini}(1993)}]{Huang:1993lm}%
  \BibitemOpen
  \bibfield  {author} {\bibinfo {author} {\bibfnamefont {HM}~\bibnamefont
  {Huang}}\ and\ \bibinfo {author} {\bibfnamefont {JL}~\bibnamefont {Kokini}},\
  }\bibfield  {title} {\enquote {\bibinfo {title} {{Measurement of biaxial
  extensional viscosity of wheat-flour doughs}},}\ }\href@noop {} {\bibfield
  {journal} {\bibinfo  {journal} {Journal of Rheology}\ }\textbf {\bibinfo
  {volume} {37}},\ \bibinfo {pages} {879--891} (\bibinfo {year}
  {1993})}\BibitemShut {NoStop}%
\bibitem [{\citenamefont {Hachmann}\ and\ \citenamefont
  {Meissner}(2003)}]{Hachmann:2003kl}%
  \BibitemOpen
  \bibfield  {author} {\bibinfo {author} {\bibfnamefont {P}~\bibnamefont
  {Hachmann}}\ and\ \bibinfo {author} {\bibfnamefont {J}~\bibnamefont
  {Meissner}},\ }\bibfield  {title} {\enquote {\bibinfo {title} {{Rheometer for
  equibiaxial and planar elongations of polymer melts}},}\ }\href@noop {}
  {\bibfield  {journal} {\bibinfo  {journal} {Journal of Rheology}\ }\textbf
  {\bibinfo {volume} {47}},\ \bibinfo {pages} {989--1010} (\bibinfo {year}
  {2003})}\BibitemShut {NoStop}%
\bibitem [{\citenamefont {Cathey}\ and\ \citenamefont
  {Fuller}(1988)}]{Cathey:1988ip}%
  \BibitemOpen
  \bibfield  {author} {\bibinfo {author} {\bibfnamefont {CA}~\bibnamefont
  {Cathey}}\ and\ \bibinfo {author} {\bibfnamefont {GG}~\bibnamefont
  {Fuller}},\ }\bibfield  {title} {\enquote {\bibinfo {title} {{Uniaxial and
  biaxial extensional viscosity measurements of dilute and semi-dilute
  solutions of rigid rod polymers}},}\ }\href@noop {} {\bibfield  {journal}
  {\bibinfo  {journal} {Journal of Non-Newtonian Fluid Mechanics}\ }\textbf
  {\bibinfo {volume} {30}},\ \bibinfo {pages} {303--316} (\bibinfo {year}
  {1988})}\BibitemShut {NoStop}%
\bibitem [{\citenamefont {Walker}\ \emph {et~al.}(1996)\citenamefont {Walker},
  \citenamefont {Moldenaers},\ and\ \citenamefont {Berret}}]{Walker1996:kk}%
  \BibitemOpen
  \bibfield  {author} {\bibinfo {author} {\bibfnamefont {LM}~\bibnamefont
  {Walker}}, \bibinfo {author} {\bibfnamefont {P}~\bibnamefont {Moldenaers}}, \
  and\ \bibinfo {author} {\bibfnamefont {JF}~\bibnamefont {Berret}},\
  }\bibfield  {title} {\enquote {\bibinfo {title} {{Macroscopic response of
  wormlike micelles to elongational flow}},}\ }\href@noop {} {\bibfield
  {journal} {\bibinfo  {journal} {Langmuir}\ }\textbf {\bibinfo {volume}
  {12}},\ \bibinfo {pages} {6309--6314} (\bibinfo {year} {1996})}\BibitemShut
  {NoStop}%
\bibitem [{\citenamefont {Venerus}\ \emph {et~al.}(2019)\citenamefont
  {Venerus}, \citenamefont {Mick},\ and\ \citenamefont
  {Kashyap}}]{Venerus:2019cj}%
  \BibitemOpen
  \bibfield  {author} {\bibinfo {author} {\bibfnamefont {David~C}\ \bibnamefont
  {Venerus}}, \bibinfo {author} {\bibfnamefont {Rebecca~M}\ \bibnamefont
  {Mick}}, \ and\ \bibinfo {author} {\bibfnamefont {Teresita}\ \bibnamefont
  {Kashyap}},\ }\bibfield  {title} {\enquote {\bibinfo {title} {{Equibiaxial
  elongational rheology of entangled polystyrene melts}},}\ }\href@noop {}
  {\bibfield  {journal} {\bibinfo  {journal} {Journal of Rheology}\ }\textbf
  {\bibinfo {volume} {63}},\ \bibinfo {pages} {157--165} (\bibinfo {year}
  {2019})}\BibitemShut {NoStop}%
\bibitem [{\citenamefont {Marrucci}\ and\ \citenamefont
  {Ianniruberto}(2004)}]{Marrucci:2004cd}%
  \BibitemOpen
  \bibfield  {author} {\bibinfo {author} {\bibfnamefont {Giuseppe}\
  \bibnamefont {Marrucci}}\ and\ \bibinfo {author} {\bibfnamefont {Giovanni}\
  \bibnamefont {Ianniruberto}},\ }\bibfield  {title} {\enquote {\bibinfo
  {title} {{Interchain Pressure Effect in Extensional Flows of Entangled
  Polymer Melts}},}\ }\href@noop {} {\bibfield  {journal} {\bibinfo  {journal}
  {Macromolecules}\ }\textbf {\bibinfo {volume} {37}},\ \bibinfo {pages}
  {3934--3942} (\bibinfo {year} {2004})}\BibitemShut {NoStop}%
\bibitem [{\citenamefont {Rayleigh}(1879)}]{Rayleigh:1879vw}%
  \BibitemOpen
  \bibfield  {author} {\bibinfo {author} {\bibfnamefont {F.R.S}\ \bibnamefont
  {Rayleigh}},\ }\bibfield  {title} {\enquote {\bibinfo {title} {{On the
  capillary phenomena of jets}},}\ }\href@noop {} {\bibfield  {journal}
  {\bibinfo  {journal} {Proc. R. Soc. Lond.}\ }\textbf {\bibinfo {volume}
  {29}},\ \bibinfo {pages} {71--97} (\bibinfo {year} {1879})}\BibitemShut
  {NoStop}%
\end{thebibliography}%

\end{document}